\documentclass[aps,prd,twocolumn,amssymb,eqsecnum,showpacs,superscriptaddress,nofootinbib,floatfix]{revtex4}

\usepackage{bm}
\usepackage{amsmath}
\usepackage{euscript}
\usepackage{graphicx}

\begin{document}

\title{A census of transient orbital resonances encountered during
  binary inspiral}

\author{Uchupol Ruangsri}
\affiliation{Department of Physics and MIT Kavli Institute, MIT, 77
  Massachusetts Ave., Cambridge, MA 02139}

\author{Scott A.\ Hughes}
\affiliation{Department of Physics and MIT Kavli Institute, MIT, 77
  Massachusetts Ave., Cambridge, MA 02139}
\affiliation{Canadian Institute for Theoretical Astrohysics,
  University of Toronto, 60 St.\ George St., Toronto, ON M5S 3H8,
  Canada}
\affiliation{Perimeter Institute for Theoretical Physics, Waterloo, ON
  N2L 2Y5, Canada}

\begin{abstract}
Transient orbital resonances have recently been identified as
potentially important to the inspiral of small bodies into large black
holes.  These resonances occur as the inspiral evolves through moments
in which two fundamental orbital frequencies, $\Omega_\theta$ and
$\Omega_r$, are in a small integer ratio to one another.  Previous
work has demonstrated that a binary's parameters are ``kicked'' each
time the inspiral passes through a resonance, changing the orbit's
characteristics relative to a model that neglects resonant effects.
In this paper, we use exact Kerr geodesics coupled to an accurate but
approximate model of inspiral to survey orbital parameter space and
estimate how commonly one encounters long-lived orbital resonances.
We find that the most important resonances last for a few hundred
orbital cycles at mass ratio $10^{-6}$, and that resonances are almost
certain to occur during the time that a large mass ratio binary would
be a target of gravitational-wave observations.  Resonances appear to
be ubiquitous in large mass ratio inspiral, and to last long enough
that they are likely to affect binary evolution in observationally
important ways.
\end{abstract}
\maketitle

\section{Introduction}
\label{sec:intro}

\subsection{Motivation}

The evolution of large mass-ratio binaries has been a particularly
active focus of research in the gravity community in recent years.
This work has been motivated in part by the promise of astrophysical
extreme mass ratio inspirals (or ``EMRIs'') as important sources for
low-frequency gravitational-wave detectors, but also because this is a
limit of the general relativistic two-body problem that can in
principle be solved with high precision.  By treating the mass ratio
of the system as a small parameter, the tools of black hole
perturbation theory can be applied, treating the spacetime of the
large black hole as ``background,'' and the effects of the smaller
body as a perturbation to that background.

On short timescales, the smaller body moves on a trajectory that is
nearly a geodesic orbit of the binary's large black hole.  ``Nearly''
refers to a {\it self force} arising from the small body's interaction
with its perturbation to the spacetime that pushes the small body away
from the geodesic worldline; see {\cite{ppv,barack}} for reviews of
the self force research program and snapshots of recent progress.
This self force has a dissipative component which is responsible for
inspiral, and a conservative component which shifts the
short-timescale motion with respect to geodesics.  Because the motion
remains close to Kerr geodesic orbits in a meaningful sense, it is
useful to use geodesics as a standard against which the motion is
compared.

Bound Kerr geodesics are triperiodic {\cite{schmidt}}.  Each orbit has
a frequency $\Omega_r$ that describes radial oscillations, a frequency
$\Omega_\theta$ that describes polar oscillations, and a frequency
$\Omega_\phi$ that describes rotations about the black hole's spin
axis.  In the weak field, these three frequencies become equal to one
another, asymptoting to the frequency predicted by Kepler's law.  They
differ significantly in the strong field, with $\Omega_r$ generically
smaller than $\Omega_\theta$ and $\Omega_\phi$.  As a small body
spirals into the strong-field of a large black hole, these frequencies
evolve at different rates.

The different rates of evolution for these frequencies have
potentially important consequences for how the self interaction
affects the binary.  For a slowly evolving system, each component of
the self force can be written as a Fourier series in terms of the
underlying fundamental frequencies {\cite{dh04}}:
\begin{equation}
f^\mu = \sum_{k = -\infty}^\infty\sum_{n = -\infty}^\infty f^\mu_{kn}
e^{-i(k\Omega_\theta + n\Omega_r)t}\;.
\label{eq:fourier_schematic}
\end{equation}
(The $\phi$ frequency does not enter this expansion since, by
axisymmetry, the self force cannot depend on the axial position of the
small body.)  For a slowly evolving orbit, the Fourier components
$f^\mu_{kn}$ and the frequencies themselves slowly change as the orbit
proceeds.  For most orbits, Eq.\ (\ref{eq:fourier_schematic})
indicates that the self interaction is given by a secularly evolving
near-constant piece ($f^\mu_{00}$), plus many rapidly oscillating terms
(all terms with $k \ne 0$, $n \ne 0$).  The rapidly oscillating terms
average to zero over multiple orbits, and the bulk of the self
interaction arises from $f^\mu_{00}$.

The strong-field behavior of Kerr black hole orbits can significantly
change how the self force behaves on average.  Some Kerr orbits are
{\it commensurate}: their fundamental frequencies can be written
$\Omega_\theta /\Omega_r = \beta_\theta/\beta_r$, where $\beta_\theta$
and $\beta_r$ are small integers.  When this is the case, there will
be a set of integers $(k,n)$ such that
\begin{equation}
k\Omega_\theta + n\Omega_r = 0\;.
\label{eq:resonance}
\end{equation}
All terms in Eq.\ (\ref{eq:fourier_schematic}) for which
Eq.\ (\ref{eq:resonance}) holds will be non-oscillatory, and so will
not average away over multiple orbits.  {\it Resonant} orbits, for
which Eq.\ (\ref{eq:resonance}) holds for some $k$ and $n$, have the
potential to significantly change the evolution of binary systems near
commensurate orbits as compared to the ``normal'' adiabatic evolution.

The importance of resonances has been the focus of several recent
papers, examining their role in the astrophysics and orbital dynamics
of black holes {\cite{bgh}}, the possibility that they may allow a
significant speedup in the computation of radiation from Kerr orbits
{\cite{glg}}, as well as their impact on the self interaction of small
bodies orbiting black holes.  The first analysis looking at the self
interaction was by Hinderer and Flanagan {\cite{hf08,fh12}}, who
developed the physical picture sketched above, and studied these
systems in more detail by coupling exact Kerr geodesics to a
post-Newtonian self-force model.  They found that this resonant
behavior near commensurate orbits can significantly change the
evolution of a binary, ``kicking'' the evolution of the orbit's
conserved integrals as compared to an analysis that does not take the
resonance into effect.

More quantitatively, let ${\cal C}$ stand for one of the integrals
that characterizes Kerr geodesics (the energy $E$, the angular
momentum $L_z$, or the Carter constant $Q$).  Hinderer and Flanagan
show that near a resonance, the rate of change $\dot{\cal C}$ acquires
a non-trivial dependence on the relative phase $\chi_0$ of the
binary's $\theta$ and $r$ motions.  Evolving the system through
resonance, one finds that ${\cal C}$ is kicked by an amount
$\Delta{\cal C}$ that depends on this phase, by the amount that
$\dot{\cal C}$ differs from its average value, and by the time $T_{\rm
  res}$ that the system is ``close to'' the resonance:
\begin{equation}
\Delta{\cal C} = \epsilon \left[\dot{\cal C}(\chi_0) -
  \langle\dot{\cal C}\rangle\right]T_{\rm res}\;.
\label{eq:kickschematic}
\end{equation}
``Close to'' a resonance is defined more precisely in
Sec.\ {\ref{sec:dsf}}, and the phase $\chi_0$ is defined in
Sec.\ {\ref{sec:gengeods}}.  The parameter $\epsilon$ depends in
detail on how the self interaction evolves as the system moves through
the resonance.  The influence of these kicks in a binary's evolution
can be seen in Fig.\ 1 of Ref.\ {\cite{fh12}} (particularly the middle
panel).

\subsection{This analysis}

We now seek to go beyond the post-Newtonian approximations used in
Refs.\ {\cite{hf08,fh12}}, with the eventual goal of self consistently
evolving a large mass ratio binary through resonances using a
strong-field dissipative self force.  As the tools to do this analysis
are being developed {\cite{fhhr}}, it will be useful to have some
estimates of what we expect from this analysis.  Our goal in this
paper is to study orbital resonances using approximate tools that are
simpler to use than rigorously assembling the strong-field dissipative
self force.

Previous work {\cite{fhr}} has given some insight into how self force
components behave exactly on resonance, computing the on-resonance
rates of change $\dot{\cal C}(\chi_0)$.  Our goal now is to begin to
understand how systems behave in the vicinity of resonances.  We
examine a sequence of exact Kerr geodesic orbits, slowly evolving
through this sequence using the Gair-Glampedakis (hereafter ``GG'')
``kludge'' approximation {\cite{gg}}.  Combining strong-field
geodesics with the GG formulas allows us to compute sequences
$[\Omega_r(t), \Omega_\theta(t)$] which approximate those that
describe a small body spiraling into a Kerr black hole in general
relativity.  The GG formulas are designed so that a system's
dissipation differs from rigorously computed strong-field formulas by
no more than a few percent over the parameter range expected for
astrophysical large-mass-ratio systems.  They provide good estimates
for the time it takes for a small body to spiral into a larger black
hole, and should be good tools for surveying the ``landscape'' of
orbital resonances.

The particular goal of this paper is to understand some practical
issues regarding the importance of resonances: How often do resonances
occur in realistic inspirals?  When a resonance occurs, how long does
it last?  Given an ensemble of randomly selected EMRIs, how many will
encounter a resonance during their inspiral, and will those resonances
last long enough that they are likely to have a strong integrated
effect on the system?  Answering these questions will clarify how
important it is to understand resonances when modeling large
mass-ratio binaries.  We are particularly interested in understanding
this in the context of space-based low-frequency gravitational-wave
observations, like the proposed eLISA mission {\cite{elisa_wp}}.
eLISA has a nominal mission lifetime of two years.  As such,
resonances that occur in the final two or so years of an EMRI's
evolution are very likely to have an impact on our ability to measure
their waves with this mission.

The formal answer to the question of how often resonances occur is
deceptively simple: {\it All} EMRIs encounter an {\it infinite} number
of resonances prior to the smaller body's plunge into the large black
hole.  This is because the period associated with radial motion,
$T_r$, diverges as the separatrix between stable and unstable orbits
is approached; see Sec.\ {\ref{sec:separatrix}} for discussion.  The
period $T_\theta$ remains finite in this limit, so
$\Omega_\theta/\Omega_r \to \infty$ as the separatrix is approached.

Though correct, this formal answer is not so interesting.  The vast
majority of those orbital resonances will be very short lived and
occur in rapid succession.  What is more interesting is to understand
which resonances are long lived and likely to have a strong impact on
the inspiral.  That is the purpose of this paper.

Previous work {\cite{fhr}} has shown that the 3:2, 2:1, and 3:1
orbital resonances are especially likely to be important --- the
deviation of $\dot{\cal C}(\chi_0)$ from $\langle\dot{\cal C}\rangle$
in Eq.\ (\ref{eq:kickschematic}) can be quite large, a necessary
condition for the resonant parameter kick to likewise be large.  In
this paper, we study how often these resonances occur and how long
they last for a wide range of inspirals.  We pick a set of initial
conditions that cover a range similar to what is expected for
astrophysical EMRIs as observed by an instrument like eLISA, and then
evolve them using the GG approximation.

We find that {\it every} inspiral encounters at least one of these
resonance during the observationally important final two years of
inspiral.  Many inspirals encounter two resonances, and a few
encounter all three.  These numbers depend strongly on binary
parameters, particularly mass ratio.  Setting the large black hole to
mass $M = 10^6\,M_\odot$ and the small one to $\mu = 1\,M_\odot$, we
find that the 3:1 resonance occurs in the last two years of inspiral
for all the cases we have examined.  The 2:1 resonance occurs during
these two years if the binary's initial orbital inclination is not too
high (angle $\theta_{\rm inc}$, defined precisely in
Sec.\ {\ref{sec:gengeods}}, less than about $80^\circ$), and the
initial eccentricity is low ($e_{\rm init} \lesssim 0.4$).  At these
masses, the 3:2 resonance does not occur during the final two years
for any case we examine, although it is just slightly outside this
window for very shallow ($\theta_{\rm inc} \lesssim 10^\circ$), low
eccentricity ($e_{\rm init} \lesssim 0.4$) inspirals into rapidly
spinning ($a \gtrsim 0.7M$) black holes.  Because the inspiral time
scales as $M^2/\mu$, we can find cases in which all three resonances
occur in the final two years if $M$ is smaller than $10^6\,M_\odot$,
or $\mu$ larger than $1\,M_\odot$.

In all cases, the 3:2 resonance lasts longest, meaning that the EMRI
spends the most time near this resonance; the 3:1 resonance is the
shortest.  (``Near'' resonance is defined by examining a phase
variable which is stationary exactly on resonance.  We define a system
to be ``near'' resonance when this variable is within 1 radian of its
value exactly on the resonance.)  The fact that the 3:2 resonance
lasts so long is a simple consequence of the fact that it occurs in
the relative weak field (at the largest orbital radius of the
resonances we survey).  Conversely, the 3:1 resonance is in the
relative strong field.  Even the shortest resonance we survey lasts
for over 50 oscillations for our $10^6$:$1$ binary; most of the
resonances we examine last for several hundred oscillations at these
masses.  The number of near-resonance oscillations scales as
$\sqrt{M/\mu}$, so we generically expect many dozens of oscillations
even for mass ratios close to $10^5$ or $10^4$.

Taken together with the on-resonance rates of change presented in
Ref.\ {\cite{fhr}}, these results amplify the developing message of
the importance of the resonant self interaction: Resonances occur
frequently enough and last long enough that they are almost certain to
have an important impact on the evolution of large mass-ratio binary
systems.  Indeed, our results imply that this behavior is generic ---
one should expect every observed EMRI to encounter a resonance.  Self
consistent modeling of a resonance's impact on binary evolution will
thus be crucial for being able to observe this systems.

The remainder of this paper is organized as follows.  We begin in
Sec.\ {\ref{sec:overview}} with a brief overview of inspiral in EMRI
systems, and of the physics of orbital resonances, highlighting what
is understood today, and what remains to be clarified.  In
Sec.\ {\ref{sec:ouranalysis}}, we describe the tools and techniques
that we use here.  We briefly describe our parameterization of Kerr
orbits (Sec.\ {\ref{sec:gengeods}}) and the Gair-Glampedakis inspiral
``kludge'' (Sec.\ {\ref{sec:ggapprox}}).  In
Sec.\ {\ref{sec:scaling}}, we show how important quantities scale with
the masses of the binary's members.  This allows us to do our detailed
numerical calculations for a single set of masses, and then to
extrapolate from there.

Section {\ref{sec:results}} presents our results.  We sample a wide
range of large black hole spins, a wide range of initial eccentricity
$e$, and a wide range of initial orbit inclination $\theta_{\rm inc}$.
We pick the initial separation so that the binary is just beyond the
3:2 resonance, and then evolve with the GG approximation until the
smaller body reaches the separatrix between stable and unstable
orbits.  Figures {\ref{fig:spacing}} and {\ref{fig:duration}} show our
results for representative EMRIs; further data are given in Tables
{\ref{tab:spacinga0.1}} -- {\ref{tab:dura0.9}}.  (To keep the main
body of the paper from being swamped with data, these tables are given
in Appendices {\ref{app:timeleft}} and {\ref{app:duration}}.)

Our conclusions are given in Sec.\ {\ref{sec:conclude}}.  As already
mentioned, our main conclusion is that resonances are ubiquitous and
long lasting: We expect every astrophysical EMRI to encounter at least
one orbital resonance en route to the binary's final plunge, and for
this resonance to last long enough that it is likely to have a
substantial impact on the system's evolution.  This strongly motivates
efforts to self consistently model how resonances impact extreme mass
ratio binaries.  We wrap up this paper by sketching plans for this
analysis.

Throughout this paper, we use ``relativist's units,'' with $G = 1 =
c$.  A useful conversion in these units is $1\,M_\odot = 4.92 \times
10^{-6}$ seconds.

\section{Overview: Inspiral and resonant self interaction}
\label{sec:overview}

We begin with a brief overview of the self interaction and how it
drives inspiral in large mass ratio binaries.  We particularly
highlight how the self interaction behaves near commensurate orbits,
showing how to compute the quantities which will be the focus of the
remainder of this paper.

\subsection{Overview of inspiral}
\label{sec:inspiral}

Consider a small body moving in the spacetime of a large black hole.
The most common astrophysical motivation for this setup is the
``extreme mass ratio inspiral'' or EMRI, in which a compact body of
mass $\mu = 1-100\,M_\odot$ is captured onto a relativistic orbit of a
black hole of mass $M = 10^{5-7}\,M_\odot$.  We emphasize that this
scenario can be taken as the large mass-ratio limit of the two-body
problem in general relativity.

At zeroth order in the mass ratio, the small body follows a geodesic
of the background spacetime:
\begin{equation}
\frac{d^2z^\mu}{d\tau^2} + {\Gamma^\mu}_{\alpha\beta}
\frac{dz^\alpha}{d\tau}\frac{dz^\beta}{d\tau} = 0\;.
\label{eq:geodesic}
\end{equation}
The parameter $\tau$ is proper time along the small body's worldline,
$z^\mu(\tau)$ is its position along the worldline, and
${\Gamma^\mu}_{\alpha\beta}$ is the connection of the background
spacetime.  We will take this background to be a Kerr black hole.  The
symmetries of Kerr admit a set of conserved integrals of the motion
which allow us to write Eq.\ (\ref{eq:geodesic}) as first-order
equations describing motion in the four Boyer-Lindquist coordinates
{\cite{mtw}}:
\begin{eqnarray}
\Sigma^2\left(\frac{dr}{d\tau}\right)^2 &=& \left[E(r^2+a^2) - a
  L_z\right]^2
\nonumber\\
& & - \Delta\left[r^2 + (L_z - a E)^2 + Q\right]
\equiv R(r)\;,
\nonumber\label{eq:rdot}\\
\Sigma^2\left(\frac{d\theta}{d\tau}\right)^2 &=& Q - \cot^2\theta L_z^2
 -a^2\cos^2\theta(1 - E^2)\;,\label{eq:thetadot}\\
\Sigma\left(\frac{d\phi}{d\tau}\right) &=&
\csc^2\theta L_z + aE\left(\frac{r^2+a^2}{\Delta} - 1\right)
-\frac{a^2L_z}{\Delta}\nonumber\\
\label{eq:phidot}\\
\Sigma\left(\frac{dt}{d\tau}\right) &=&
E\left[\frac{(r^2+a^2)^2}{\Delta} - a^2\sin^2\theta\right]
\nonumber\\
& & + aL_z\left(1 - \frac{r^2+a^2}{\Delta}\right)
\;.\label{eq:tdot}
\end{eqnarray}
In these equations, $\Sigma = r^2 + a^2\cos^2\theta$, and $\Delta =
r^2 - 2Mr + a^2$.  The quantities $E$ and $L_z$ are the orbital energy
and axial angular momentum, normalized to the mass $\mu$ of the
orbiting body.  They are related to the Kerr spacetime's timelike and
axial Killing vectors,
\begin{eqnarray}
E &=& -\xi_\mu^{(t)}u^\mu\;,
\label{eq:Edef}\\
L_z &=& \xi_\mu^{(\phi)}u^\mu\;,
\label{eq:Lzdef}
\end{eqnarray}
where $u^\mu \equiv dz^\mu/d\tau$ is the 4-velocity of the small body.
The quantity $Q$ is the orbit's Carter constant, normalized to
$\mu^2$.  It is related to a Killing tensor admitted by the Kerr
spacetime:
\begin{equation}
Q = Q_{\mu\nu} u^\mu u^\nu\;.
\label{eq:Qdef}
\end{equation}
These three quantities are conserved on any geodesic.

As described in the introduction, bound Kerr geodesics and any
function which arises from them can be described in the frequency
domain {\cite{schmidt,dh04}} in terms of three fundamental orbital
frequencies.  Consider some function ${\cal F}$ that depends on a
bound Kerr geodesic worldline $z^\mu$.  Then, we can write
\begin{equation}
{\cal F}(z^\mu) = \sum_{mkn} {\cal F}_{mkn} e^{-i(m\Omega_\phi +
  k\Omega_\theta + n\Omega_r)t}\;,
\label{eq:t_expansion}
\end{equation}
where $t$ is Boyer-Lindquist time.  (Many functions, including the
self force components we will discuss below, have no dependence on
$\phi$.  For them, the $m$ index can be neglected.)  This expansion is
very useful for describing observables such as the gravitational waves
generated by a binary, since $t$ corresponds to time as measured on
the clocks of distant observers.  One can also define a
frequency-domain expansion like Eq.\ (\ref{eq:t_expansion}) using a
time parameter that is better suited to studying the strong-field
motion than $t$.  A common choice is ``Mino time,'' which decouples
the radial and polar oscillations; see {\cite{dh04}} for further
discussion of this time parameter and the frequencies associated with
it.  We will use the Boyer-Lindquist time expansion
(\ref{eq:t_expansion}) for this analysis.

At first order in the mass ratio, the spacetime is deformed by the
smaller body's mass.  The self interaction of the small body with its
own spacetime deformation can be regarded as a {\it self
  force}\footnote{The tilde on $\tilde f^\mu \equiv du^\mu/d\tau$ in
  Eq.\ (\ref{eq:forcedgeodesic}) indicates that this describes the
  self interaction per unit proper time.  We leave the tilde off
  elsewhere in this paper, indicating the self interaction per unit
  coordinate time: $f^\mu \equiv du^\mu/dt$.} (or self acceleration)
pushing the small body away from the geodesic worldline:
\begin{equation}
\frac{d^2z^\mu}{d\tau^2} + {\Gamma^\mu}_{\alpha\beta}
\frac{dz^\alpha}{d\tau}\frac{dz^\beta}{d\tau} = \tilde f^\mu\;.
\label{eq:forcedgeodesic}
\end{equation}
Excellent discussion of the gravitational self force research program
and recent progress can be found in Refs.\ {\cite{ppv,barack}}.  For
our purposes, it is enough to note that the self force can be broken
into a time-symmetric conservative piece, and a time-asymmetric
dissipative piece:
\begin{equation}
f^\mu = f^\mu_{\rm diss} + f^\mu_{\rm cons}\;.
\end{equation}
On average, the dissipative self force takes energy and angular
momentum from the binary, driving the small body's inspiral, and the
conservative self force shifts the orbital frequencies from their
background values.

In this analysis, we neglect the conservative self force, focusing on
the dissipative self interaction.  We partially justify this by noting
that the integrated conservative self interaction is smaller than the
integrated effect of the dissipative piece by roughly the system's
mass ratio {\cite{hf08}}.  We emphasize that are not arguing that it
is correct to neglect the conservative self force (indeed, it is well
known that the conservative self force has an important impact on the
phasing of large mass ratio binaries {\cite{ppn}}), but are just
suggesting that focusing on dissipation will give insight into the
astrophysical relevance of resonances.

We further justify neglecting the conservative self force for the
pragmatic reason that the dissipative self force is presently better
understood than the conservative one.  Mature codes exist to compute
the averaged effect of dissipation {\cite{dh06}}, and it appears to be
straightforward to generalize these codes to analyze the instantaneous
dissipative self force {\cite{fhhr}}.  Detailed conservative self
forces have been computed for arbitrary orbits in the Schwarzschild
limit {\cite{bs10}}, and progress on the Kerr case is proceding apace.
It has already been noted using post-Newtonian techniques that the
conservative self interaction has a non-negligible impact on the
analysis of resonances {\cite{fh12,isoyama}}.  Flanagan and Hinderer
in particular find for at least one example that the conservative
effect, though smaller than the leading dissipative effect by about an
order of magnitude, tends to augment the dissipative effects (see
lower panel of Fig.\ 1 of Ref.\ {\cite{fh12}}).  It is not clear,
though, if this is a general feature of the self force.  The
conservative and dissipative contributions to the self force will
depend on various phases in different ways, so the augmentation that
they find may be an accident of the case they focus on for their
analysis.  It will be valuable and interesting to include strong field
conservative self forces in an analysis of this sort at some point in
the future.

\subsection{Form of the dissipative self force and its near
resonant behavior}
\label{sec:dsf}

The dissipative self force has three components, which can be regarded
as the instantaneous rate of change of the conserved integrals:
\begin{eqnarray}
\frac{dE}{dt} &=& -\xi^{(t)}_\mu f_{\rm diss}^\mu\;,
\\
\frac{dL_z}{dt} &=& \xi^{(\phi)}_\mu f_{\rm diss}^\mu\;,
\\
\frac{dQ}{dt} &=& 2Q_{\mu\nu} u^\nu f_{\rm diss}^\mu\;.
\end{eqnarray}
These components can be written
\begin{equation}
\frac{d{\cal C}}{dt} = \dot{\cal C}_{00} +
\sum_{\substack{k=-\infty\\k\ne0}}^\infty
\sum_{\substack{n=-\infty\\n\ne0}}^\infty
\dot{\cal C}_{kn} e^{-i\Phi_{kn}(t)}\;,
\label{eq:dsfcomponent}
\end{equation}
where ${\cal C}$ stands for $E$, $L_z$, or $Q$.  We separate the $00$
components to emphasize that these components survive any long-time
averaging; we discuss this further below.  Details of how to compute
the coefficients $\dot{\cal C}_{kn}$ will be given in a forthcoming
paper {\cite{fhhr}}; an example for the self force acting on a scalar
charge is given in Ref.\ {\cite{dfhcqg}}.  The phase variable
appearing in Eq.\ (\ref{eq:dsfcomponent}) is given by
\begin{equation}
\Phi_{kn}(t) = \int_{t_0}^t \left[k\Omega_\theta(t') +
  n\Omega_r(t')\right]dt'\;,
\label{eq:dsfphase}
\end{equation}
where $t_0$ is a starting time, defining when measurements of the
system begin.

For most times and most index pairs $(k,n)$, the $e^{i\Phi_{kn}(t)}$
factor in Eq.\ (\ref{eq:dsfphase}) oscillates rapidly.  When this is
the case, then on average only the $00$ term contributes to $d{\cal
  C}/dt$.  However, if it happens that the polar and radial
frequencies are {\it commensurate}, meaning $\Omega_\theta/\Omega_r =
\beta_\theta/\beta_r$ with $\beta_\theta$ and $\beta_r$ both integers,
then the situation can change significantly.  When $k$ and $n$ satisfy
$k\beta_\theta + n\beta_r = 0$, then the phase will be {\it
  stationary} near that time, and the exponential factor in
Eq.\ (\ref{eq:dsfphase}) will vary very slowly.

\begin{figure}[ht]
\includegraphics[width = 0.48\textwidth]{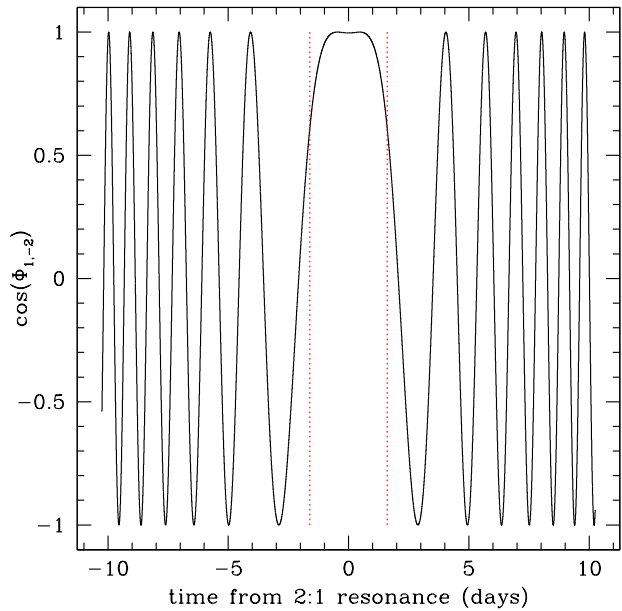}
\caption{Behavior of $\cos\Phi_{kn}(t)$ for $k = 1$, $n = -2$ near a
  moment in inspiral at which $\Omega_\theta/\Omega_r = 2$.  To
  generate this plot, we take the large black hole to have spin $a =
  0.7M$ and mass $M = 10^6\,M_\odot$; the small body has mass $\mu =
  1\,M_\odot$.  The inspiral has starting parameters $p = 7.909M$, $e
  = 0.4$, and $\theta_{\rm inc} = 30^\circ$.  Further data related to
  this case is presented in Fig.\ {\ref{fig:binevolve}} and Tables
  {\ref{tab:spacinga0.7}} and {\ref{tab:dura0.7}}.}
\label{fig:statphase}
\end{figure}

This behavior is illustrated in Fig.\ {\ref{fig:statphase}}.  We show
$\cos\Phi_{kn}(t)$ for an example inspiral near a moment when
$\Omega_\theta/\Omega_r = 2$; we describe how this inspiral is
computed in more detail below.  We have put $k = 1$ and $n = -2$, and
have shifted the time axis so that $t = 0$ corresponds to the moment
when $\Omega_\theta/\Omega_r = 2$.  This function oscillates rapidly
before and after encountering these commensurate frequencies, but
varies much more slowly at times near $t = 0$.  The vertical red lines
mark when $\Phi_{kn}(t)$ differs by 1 radian from its value at $t =
0$.  The function oscillates rapidly outside of this window, but
evolves very slowly inside.

With this behavior in mind, let us expand $\Phi_{kn}(t)$ in a Taylor
series.  Let $t_{\rm res}$ be the time at which the system encounters
the resonance, and expand $\Phi_{kn}$:
\begin{eqnarray}
\Phi_{kn}(t) &=& \Phi_{kn}(t_{\rm res}) +
(k\Omega_\theta + n\Omega_r)(t - t_{\rm res})
\nonumber\\
& & + \frac{1}{2} (k\dot\Omega_\theta + n\dot\Omega_r)(t - t_{\rm res})^2
+ \ldots
\nonumber\\
&\simeq& \Phi_{kn}(t_{\rm res}) + \frac{1}{2} (k\dot\Omega_\theta +
n\dot\Omega_r)(t - t_{\rm res})^2\;.
\nonumber\\
\label{eq:Phi_kn_taylor}
\end{eqnarray}
The frequencies and frequency derivatives here are all evaluated at
$t_{\rm res}$, and $(k,n)$ are chosen so that $k\Omega_\theta +
n\Omega_r = 0$ at $t = t_{\rm res}$.  Let us define the system as
being ``near'' resonance when $\Phi_{kn}$ differs from
$\Phi_{kn}(t_{\rm res})$ by no more than one radian.  By
Eq.\ (\ref{eq:Phi_kn_taylor}), the amount of time spent near resonance
is then given by
\begin{equation}
T_{\rm res} = 2\sqrt{\frac{2}{k\dot\Omega_\theta(t_{\rm res}) +
    n\dot\Omega_r(t_{\rm res})}}\;.
\label{eq:Tres}
\end{equation}
(The prefactor of $2$ means the resonance lasts from $-T_{\rm res}/2$
to $T_{\rm res}/2$.)  For the example shown in
Fig.\ {\ref{fig:statphase}}, Eq.\ (\ref{eq:Tres}) predicts $T_{\rm
  res} = 61170M$, within a few percent of the result $63140M$ found by
directly computing the times at which the phase differs by 1 radian
from $\Phi_{kn}(t_{\rm res})$.

Multiplying $T_{\rm res}$ by $\Omega_{\theta,r}$, we calculate the
number of orbits the system spends near resonance:
\begin{equation}
N_{\theta,r} = \Omega_{\theta,r} T_{\rm res}/2\pi\;.
\label{eq:Nthr}
\end{equation}
We will use $N_\theta$ and $N_r$ to describe the duration of each
resonance.

\section{Tools and techniques for our analysis}
\label{sec:ouranalysis}

Here we describe the tools we use for this analysis.  We discuss
generic bound geodesics in Sec.\ {\ref{sec:gengeods}}, and the
Gair-Glampedakis formulas we use to evolve through a sequence of these
geodesics in Sec.\ {\ref{sec:ggapprox}}.  To reduce the number of
cases we need to separately consider, we will do all of our
computations for a single set of masses ($M = 10^6\,M_\odot$, $\mu =
1\,M_\odot$), and use the scaling laws developed in
Sec.\ {\ref{sec:scaling}} to extrapolate to other masses.  Section
{\ref{sec:separatrix}} wraps up this section by discussing the
separatrix between stable and unstable orbits, and the behavior of
$\Omega_\theta/\Omega_r$ as the separatrix is approached.

\subsection{Generic bound geodesic orbits}
\label{sec:gengeods}

The geodesic equations which describe motion at any moment have been
presented in Eqs.\ (\ref{eq:rdot})--(\ref{eq:tdot}).  The key point
for our analysis is that, up to initial conditions, geodesic motion is
characterized by the three integrals $E$, $L_z$, and $Q$.  Here, we
describe in more detail how we parameterize a given orbit.  We remap
an orbit's $r$ and $\theta$ coordinates to parameters $p$, $e$, and
$\theta_{\rm m}$, defined by
\begin{eqnarray}
r &=& \frac{pM}{1 + e\cos\psi}\;,
\label{eq:p_and_e_def}
\nonumber\\
\cos\theta &=& \cos\theta_{\rm m}\cos(\chi + \chi_0)\;.
\label{eq:thetam_def}
\end{eqnarray}
Strictly speaking, we should include an offset phase $\psi_0$ in our
equation for $r$.  For geodesics, setting $\psi_0 = 0$ amounts to
choosing the time origin when the orbit passes through periapsis.  For
much of our analysis, we use the angle $\theta_{\rm inc}$ introduced
in Ref.\ {\cite{dh06}}, related to $\theta_{\rm m}$ by
\begin{equation}
\theta_{\rm inc} = \pi/2 - {\rm sgn}\left(L_z\right)\theta_{\rm m}\;.
\end{equation}
Varying $\theta_{\rm inc}$ from $0$ to $\pi$ continuously varies the
orbit from prograde equatorial to retrograde equatorial.  Any orbit
with $\theta_{\rm inc} < 90^\circ$ is ``prograde'' ($L_z$ parallel to
the black hole's spin); any orbit with $\theta_{\rm inc} > 90^\circ$
is ``retrograde'' ($L_z$ antiparallel to the spin).  Once the three
parameters $p$, $e$, and $\theta_{\rm inc}$ are known, it is
straightforward to compute the three integrals $E$, $L_z$, and $Q$.
Formulas for computing ($E$, $L_z$, $Q$) given ($p$, $e$, $\theta_{\rm
  inc}$) can be found in Ref.\ {\cite{schmidt}}.

References {\cite{schmidt,dh04}} also provide formulas describing how
to compute the three frequencies $\Omega_r$, $\Omega_\theta$, and
$\Omega_\phi$ given $p$, $e$, and $\theta_{\rm inc}$.  This means
that, for our purposes, the parameters $p$, $e$, and $\theta_{\rm
  inc}$ completely characterize Kerr black hole orbits: Once those
three parameters are fixed, all other quantities describing orbits can
be computed.  In Ref.\ {\cite{fhr}}, we showed that the offset phase
$\chi_0$ is important on resonance.  However, the self interaction
calculations we do here are cruder than those done in
Ref.\ {\cite{fhr}}, so we are not sensitive to $\chi_0$ in this
analysis.

\subsection{Evolving generic generics}
\label{sec:ggapprox}

As already mentioned, we eventually would like to self-consistently
compute the motion of the small body using a rigorously computed
dissipative self force.  We have begun developing code to compute the
instantaneous dissipative self force components $d{\cal C}/dt$
[cf.\ Eq.\ (\ref{eq:dsfcomponent})], and plan to couple this to a
prescription for computing forced motion near Kerr black holes
{\cite{gfdhb}}.  This work will be reported later {\cite{fhhr}}.

Our goals here are more modest, so a less accurate approach to
modeling the binary's evolution will be adequate.  Our main goal is
simply to understand how often a given inspiral encounters an orbital
resonance, and how long each resonance lasts, using
Eq.\ (\ref{eq:Tres}) to estimate this duration.  For this, we need to
construct the sequence of frequencies $[\Omega_\theta(t),
  \Omega_r(t)]$ describing an inspiral.  This can be easily done
provided we likewise construct the sequence $[p(t), e(t), \theta_{\rm
    inc}(t)]$, or equivalently the sequence $[E(t), L_z(t), Q(t)]$.

We use the Gair-Glampedakis (GG) ``kludge'' for evolving generic
geodesic orbits to construct these sequences.  GG provides formulas
for estimating the rate at which $E$, $L_z$, and $Q$ change due to
gravitational-wave emission.  By fitting to numerical results from
black hole perturbation theory in the limit of zero eccentricity and
zero inclination, and then using post-Newtonian results to guide their
functional form, they derive formulas which fit a wide range of
numerical results from black hole perturbation theory for arbitrary
black hole spin, large eccentricity, and large inclination.  For
example, the GG formula for the rate of change of $L_z$ is
\begin{eqnarray}
\frac{dL_z}{dt}\biggr|_{\rm GG} &=& (1 - e^2)^{3/2}\left[(1-e^2)^{-3/2}
(\dot L_z)_{\rm 2PN}(p,e,\iota,a)
\right.
\nonumber\\
& & \left.
- (\dot L_z)_{\rm 2PN}(p,0,\iota,a) + (\dot L_z)_{\rm circ-fit}\right]\;.
\label{eq:Lzdotkludge}
\end{eqnarray}
This is Eq.\ (59) of Ref.\ {\cite{gg}}.  The parameter $\iota$ is an
alternate orbital inclination angle, defined by $\cos\iota =
L_z/\sqrt{L_z^2 + Q}$.  For Schwarzschild black holes, $\iota =
\theta_{\rm inc}$; even for maximal Kerr black holes, $\iota$ only
differs from $\theta_{\rm inc}$ by a few degrees.  The function $(\dot
L_z)_{\rm 2PN}$ is a slightly modified second post-Newtownian
expression for the flux of $L_z$ carried by gravitational waves; the
function $(\dot L_z)_{\rm circ-fit}$ is a (rather complicated) fit to
black hole perturbation theory flux data for $\dot L_z$ in the limit
$e = 0$.  See Secs.\ V and VI of Ref.\ {\cite{gg}} [especially
  Eqs.\ (45) and (57)] for details.  GG provide similar formulas to
estimate $dE/dt$ and $dQ/dt$.

Reference {\cite{gg}} demonstrates that the GG approximation provides
a surprisingly good model for the inspiral of small bodies into Kerr
black holes.  As such, this approach is very well suited for our goal
of taking a census of orbital resonances encountered during inspiral.
Discussion of the resulting inspirals we make using the GG
approximation is given in Sec.\ {\ref{sec:results}}.

\subsection{Scalings with masses}
\label{sec:scaling}

One very helpful result which is independent of our use of the GG
approximation is the leading scaling of certain important quantities
with the binary's masses.  Consider the orbit's energy $E$, axial
angular momentum $L_z$, and Carter constant $Q$.  Recalling that we
use units in which $G = c = 1$, dimensionless analogs of these
quantities can be found by dividing out powers of $\mu$ and $M$:
\begin{equation}
\hat E = \frac{E}{\mu}\;,\qquad \hat L_z = \frac{L_z}{\mu M}\;,
\qquad \hat Q = \frac{Q}{\mu^2M^2}\;.
\label{eq:dimensionlessconsts}
\end{equation}
We can similarly construct dimensionless versions of the time variable
$t$ and any orbital frequency $\Omega_x$:
\begin{equation}
\hat t = t/M\;,\qquad \hat\Omega_x = M\Omega_x\;.
\label{eq:dimensionlesstime}
\end{equation}
The leading rates of change of $E$, $L_z$, and $Q$ have the following
scalings with $\mu$ and $M$:
\begin{eqnarray}
\frac{dE}{dt} &\propto& \frac{\mu^2}{M^2}\;,
\label{eq:dEdtscaling}\\
\frac{dL_z}{dt} &\propto& \frac{\mu^2}{M}\;,
\label{eq:dLzdtscaling}\\
\frac{dQ}{dt} &\propto& \mu^3\;.
\label{eq:dQdtscaling}
\end{eqnarray}
Combining Eqs.\ (\ref{eq:dEdtscaling})--(\ref{eq:dQdtscaling}) with
Eqs.\ (\ref{eq:dimensionlessconsts}) and (\ref{eq:dimensionlesstime}),
we find
\begin{equation}
\frac{d\hat{\cal C}}{d\hat t} \propto \frac{\mu}{M}
\end{equation}
for $\hat{\cal C} = \hat E$, $\hat L_z$, and $\hat Q$.  From this, we
deduce how the inspiral time scales with masses: $T_{\rm insp} \sim
{\cal C}/(d{\cal C}/dt)$, so
\begin{equation}
T_{\rm insp} \propto \frac{M^2}{\mu}\;.
\label{eq:Tinspscaling}
\end{equation}

Next we would like to understand how the duration of a resonance
scales with mass ratio.  Here we need the derivatives
$d\Omega_{\theta,r}/dt$.  The derivative of any frequency
$\hat\Omega_x$ with $\hat t$ has a very simple scaling:
\begin{eqnarray}
\frac{d\hat\Omega_x}{d\hat t} &=& 
\frac{\partial\hat\Omega_x}{\partial\hat E}\frac{d\hat E}{d\hat t} +
\frac{\partial\hat\Omega_x}{\partial\hat L_z}\frac{d\hat L_z}{d\hat t} +
\frac{\partial\hat\Omega_x}{\partial\hat Q}\frac{d\hat Q}{d\hat t}
\\
&\propto& \frac{\mu}{M}\;.
\end{eqnarray}
The second line follows from the fact that $\partial
\hat\Omega_x/\partial \hat{\cal C}$ has no dependence on $M$ or $\mu$
for $\hat{\cal C} = \hat E,\hat L_z,\hat Q$ --- these dependences have
already been scaled out at leading order.  Putting factors of $M$ and
$\mu$ back in, we see that
\begin{equation}
\frac{d\Omega_x}{dt} \propto \frac{\mu}{M^3}\;.
\end{equation}
By Eq.\ (\ref{eq:Tres}), this means that
\begin{equation}
T_{\rm res} \propto \sqrt{\frac{M^3}{\mu}}\;,
\label{eq:Tresscaling}
\end{equation}
and
\begin{equation}
N_{\theta,r} \propto \sqrt{\frac{M}{\mu}}\;.
\label{eq:Nthrscaling}
\end{equation}
These scalings should be accurate over the extreme mass-ratio regime.
We will therefore do our calculations for only a single pair of masses
($\mu = 1\,M_\odot$, $M = 10^6\,M_\odot$), and extrapolate from that
using these results.

\begin{figure}[ht]
\includegraphics[width = 0.48\textwidth]{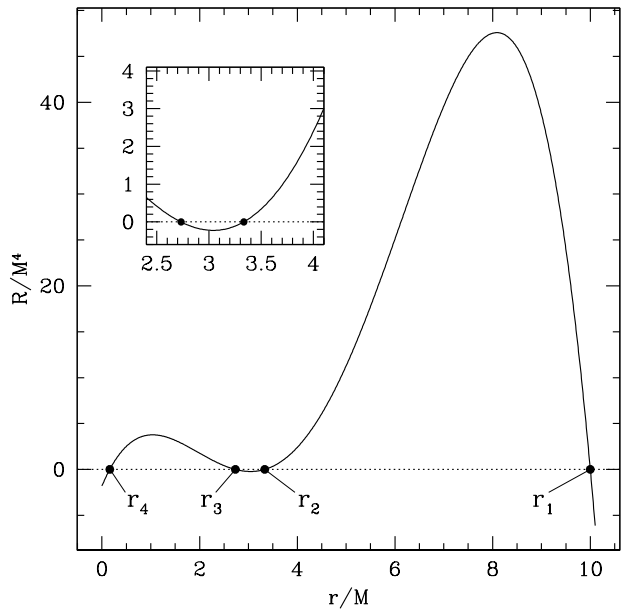}
\caption{Behavior of $R(r)$ for a stable orbit.  This example is for
  an orbit about a black hole with $a = 0.7M$; it has $p = 5M$ and $e
  = 0.5$ (so motion is confined to $3.333M \le r \le 10M$), and has
  inclination $\theta_{\rm inc} = 40^\circ$.  The root $r_3 \simeq
  2.73M$; $r_1$ through $r_4$ are indicated on the plot).  As $p$,
  $e$, and $a$ are held fixed and $\theta_{\rm inc}$ increased, $r_3$
  moves toward the orbit's periapsis, coinciding when $\theta_{\rm
    inc} = 54.11^\circ$.  The merging of the roots $r_2$ and $r_3$
  defines the separatrix between stable and unstable orbits.}
\label{fig:stable}
\end{figure}

\subsection{The stable/unstable separatrix}
\label{sec:separatrix}

Coupling the GG formulas to Kerr geodesics allows us to evolve orbits
under gravitational-wave emission.  This evolution drives orbits
deeper into the strong field until eventually they encounter a
separatrix between stable and unstable.  The inspiraling body then
rapidly plunges into the massive black hole.  Perez-Giz and Levin
{\cite{pgl}} provide excellent discussion of the properties of this
separatrix (noting that it can be understood by its equivalence to the
family of Kerr spacetime homoclinic orbits), including easy-to-use
formulas for computing its location in the space of bound orbits; many
useful and relevant formulas can also be found in
{\cite{fujitahikida}}.  Here we briefly discuss a few critical
properties of the separatrix as they pertain to our analysis.

Many of the properties of Kerr black hole orbits are determined by the
function $R(r)$ defined in Eq.\ (\ref{eq:rdot}).  It is a quartic
function of Boyer-Lindquist radius $r$, and can be rewritten
\begin{equation}
R(r) = (1 - E^2)(r_1 - r)(r - r_2)(r - r_3)(r - r_4)
\label{eq:potentialR}
\end{equation}
using the fact that $E < 1$ for bound orbits.  The roots $r_{1,2,3,4}$
are ordered such that $r_1 \ge r_2 \ge r_3 > r_4$.  They are
determined by $E$, $L_z$, $Q$, and $a$;
Refs.\ {\cite{pgl,fujitahikida}} provides explicit formulas for their
values.  Bound orbits oscillate between $r_1$ (apoapsis) and $r_2$
(periapsis), with $r_1 = r_2$ in the circular limit ($e = 0$).  The
root $r_4$ is typically inside the horizon ($r_4 = 0$ when $Q = 0$ or
$a = 0$), and is never reached by any bound orbit.

Orbital stability is determined by the root $r_3$.  Stable bound
orbits have $r_3 < r_2$; the separatrix is defined by the condition
$r_3 = r_2$.  Figure {\ref{fig:stable}} shows $R(r)$ for one example
of a stable orbit.  As orbits approach the separatrix, $r_3$ and $r_2$
move together, coinciding in the limit.

Along with terminating the inspiral, the separatrix has another
important property for our analysis: The radial period diverges as it
is approached.  This is simple to understand by analyzing the radial
motion using ``Mino time'' $\lambda$, the time variable which
separates the $r$ and $\theta$ motions.  Using $\lambda$ rather than
proper time $\tau$ as the independent parameter, the radial motion is
governed by the equation
\begin{equation}
\left(\frac{dr}{d\lambda}\right)^2 = {R(r)}\;.
\end{equation}
The Mino-time radial period is then
\begin{eqnarray}
\Lambda_r &=& 2\int_{r_2}^{r_1}\frac{dr}{\sqrt{R(r)}}
\nonumber\\
&=& 2\int_{r_2}^{r_1}\!\frac{dr}{\sqrt{(1 - E^2)(r_1 - r)(r - r_2)(r -
    r_3)(r - r_4)}}\;.
\nonumber\\
\label{eq:Lambda_r}
\end{eqnarray}
The Boyer-Lindquist time radial period $T_r$ is easily computed from
this; see Ref.\ {\cite{dh04}} for details.  The key point is that $T_r
\propto \Lambda_r$, and that the proportionality remains well behaved
as the separatrix is approached.

At the marginally stable orbit, Eq.\ (\ref{eq:Lambda_r}) becomes
\begin{equation}
\Lambda_r = 2\int_{r_2}^{r_1}\frac{dr}{(r - r_2)\sqrt{(1 - E^2)
(r_1 - r)(r - r_4)}}\;.
\end{equation}
This integral, and hence $\Lambda_r$, diverges.  An orbit that is
precisely on the separatrix will ``whirl'' for all time on a circular
trajectory at the orbit's apoapsis --- the hallmark of homoclinic
orbits.  By their proportionality, it follows that $T_r$ also diverges
as the separatrix is approached.  A similar analysis shows that
$T_\theta$ does {\it not} diverge at the separatrix, but just takes
some finite value.  The ratio $\Omega_\theta/\Omega_r$ therefore
diverges on the separatrix, proving that (formally, at least)
inspiraling bodies pass through an infinite number of orbital
resonances en route to their final plunge into the large black hole.
In practice, the small body will transition to a plunging trajectory
near the separatrix.

\begin{figure*}[ht]
\includegraphics[width = 0.48\textwidth]{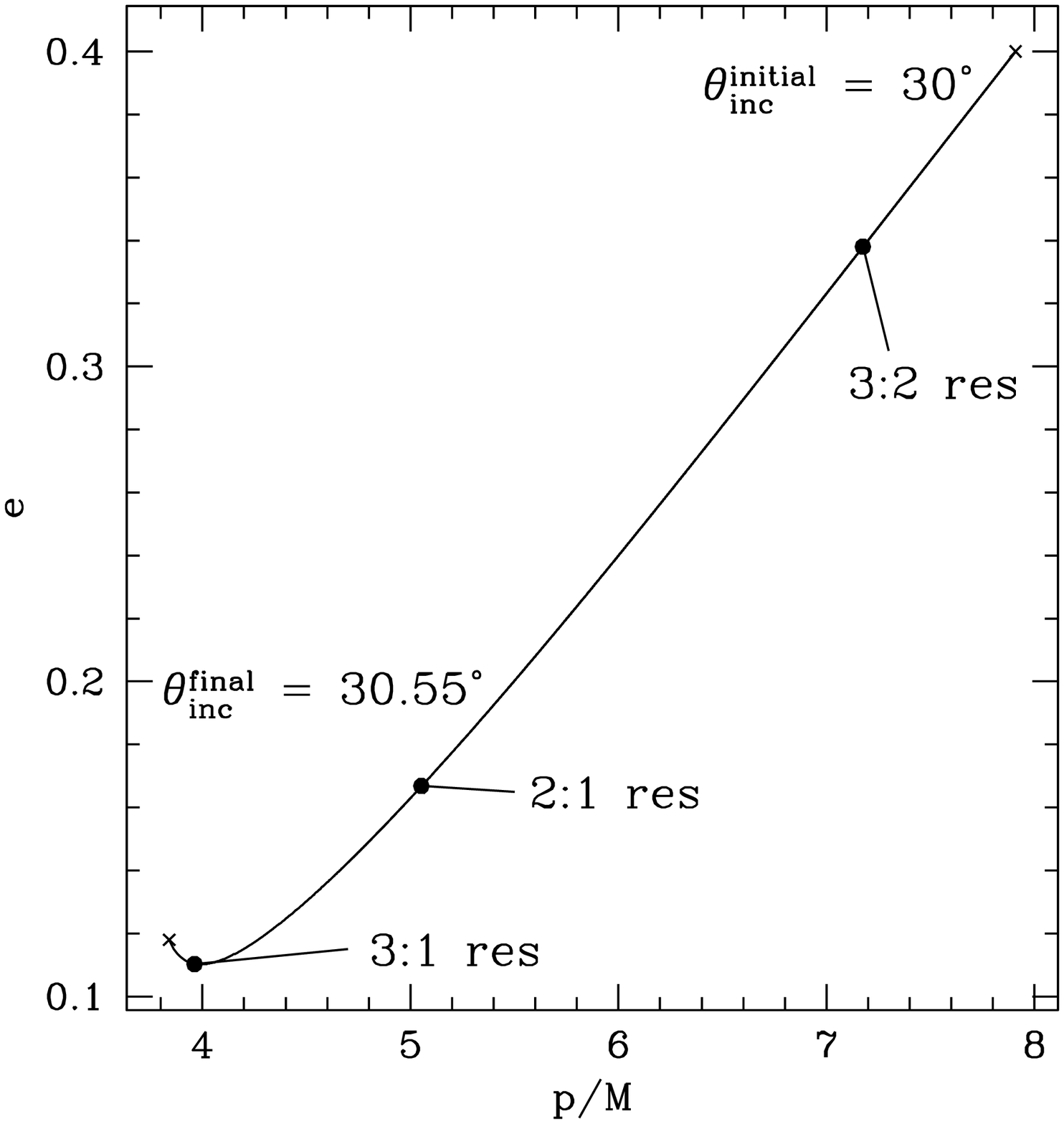}
\includegraphics[width = 0.48\textwidth]{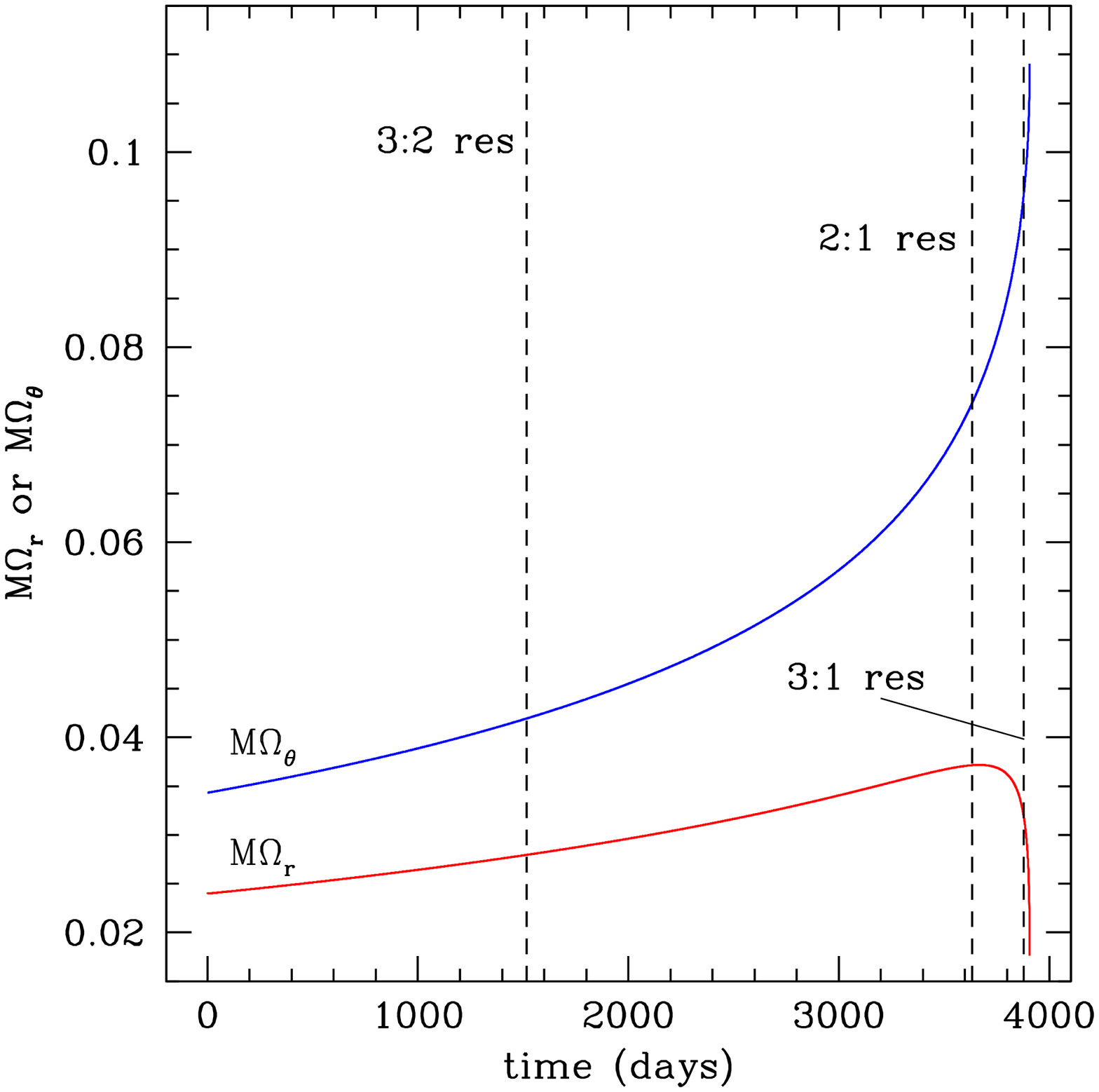}
\caption{Behavior of important quantities describing an inspiral in
  the GG approximation; the case shown here is the same one used to
  generate Fig.\ {\ref{fig:statphase}}.  Left panel shows the
  evolution of $p$, $e$, and $\theta_{\rm inc}$.  Inspiral starts at
  the upper right cross, and ends at the lower left cross.  During
  evolution, $p$ shrinks until the system reaches the separatrix; $e$
  mostly shrinks over inspiral, growing for a short time near the end.
  The inclination angle $\theta_{\rm inc}$ is nearly constant over
  inspiral, increasing by about half a degree.  The three dots on the
  track mark the moments that the binary passes through the resonances
  we study.  Right panel shows the evolution of $\Omega_r$ and
  $\Omega_\theta$ as functions of time.  Dashed vertical lines mark
  the resonances.  We have done these calculations with masses $M =
  10^6\,M_\odot$, $\mu = 1\,M_\odot$.  To extrapolate to other masses,
  use the fact that the time axis scales as $M^2/\mu$, and that all
  other quantities shown here are invariant to mass scaling.}
\label{fig:binevolve}
\end{figure*}

\section{Results}
\label{sec:results}

We now examine some key properties of extreme mass ratio systems as
they evolve through resonances.  We focus on two aspects of the
system: How much inspiral remains after passing through a resonance;
and how many oscillations in $\theta$ and $r$ are executed near
resonance.

The time remaining following each resonance is an indicator of how
likely the resonance is to be of observational importance.  As
discussed in the Introduction, the most observationally important EMRI
systems are those which are a few months or years away from their
final plunge.  A space-based low-frequency gravitational-wave detector
like eLISA will be able to track the evolution of EMRI waves during
these last few months or years.  A resonance that occurs during this
window is very likely to have direct observational consequences.
Templates describing these systems will need to account for how the
resonance modifies the binary's ``normal'' adiabatic evolution in
order to remain in phase with the astrophysical waveform.

The number of oscillations executed at each resonance is directly
related to the ``kick'' imparted to the system as it evolves through
resonance.  We emphasize again that to build a full picture of the
resonance's impact we must self-consistently integrate the system
under the influence of the dissipative self force, which we plan to do
in future work {\cite{fhhr}}.  However, in the absence of this
self-consistent picture, we can say with confidence that longer
resonances are likely to imply greater ``kicks'' to orbital
parameters, and vice versa.  The duration of a resonance is thus an
important indicator of its possible importance in the evolution of an
EMRI.

Before discussing our results in detail, we first describe how we
select the different inspirals that we study.  Our goal is to cover a
range of parameters which span what is likely for astrophysical EMRI
events that may be measured by a future low-frequency GW mission,
while keeping the total number of cases we study reasonable.  The
algorithm we follow is as follows:

\begin{itemize}

\item Choose the large black hole's spin from the set $a/M \in [0.1,
  0.4, 0.7, 0.9]$.  This range nicely covers the possibilities from
  slow to rapid spin.

\item Pick initial orbital eccentricity from the set $e \in [0.1, 0.4,
  0.7, 0.95]$.  This set covers a large portion of the range that is
  likely to describe astrophysical EMRI events {\cite{gairetal}}.

\item Pick initial orbital inclination from the set $\theta_{\rm inc}
  \in [10^\circ, 30^\circ, 60^\circ, 85^\circ, 95^\circ, 120^\circ,
    150^\circ, 170^\circ]$.  This set surveys a range from nearly
  prograde equatorial orbits to nearly retrograde equatorial orbits.
  We are careful not to examine orbits precisely at $90^\circ$, since
  the GG approximation may not be reliable there.  Because
  $\theta_{\rm inc}$ evolves only slightly {\cite{h2000,gg}}, its
  value is nearly fixed to its initial value over inspiral.

\item Compute the value of $p$ so that an orbit is exactly on the 3:2
  resonance.  We then inflate that value by $10\%$, and allow the
  binary to evolve using the GG approximation
  (Sec.\ {\ref{sec:ggapprox}}).

\end{itemize}

\noindent
In all cases, we fix the binary's masses to $M = 10^6\,M_\odot$, $\mu
= 1\,M_\odot$, and extrapolate to other masses using the results of
Sec.\ {\ref{sec:scaling}}.

Figure {\ref{fig:binevolve}} shows an example binary selected by this
procedure and then evolved with the GG approximation.  The large black
hole has $a = 0.7M$, and the binary begins with $e = 0.4$,
$\theta_{\rm inc} = 30^\circ$.  (These are the parameters used to
illustrate the near resonance stationary phase of the self interaction
in Fig.\ {\ref{fig:statphase}}.)  Left panel shows how the binary
evolves across the $(p,e)$ plane --- shrinking in $p$ over the entire
inspiral, shrinking in $e$ until very late in the inspiral, and with
$\theta_{\rm inc}$ nearly constant.  Notice that as the binary crosses
the three orbital resonances that we study, $p$ and $e$ are smaller
(in some cases, significantly smaller) than their starting values.
The right panel shows how the binary's frequencies $\Omega_\theta$ and
$\Omega_r$ evolve with time, with vertical dashed lines labeling the
moments that the system passes through orbital resonances.

Although differing in detail, all of the examples we study are
qualitatively similar to this case.  In particular, we always find
that $p$ shrinks over the entire inspiral; $e$ shrinks until the very
end, at which point it grows slightly.  (This behavior was originally
discovered in studies of eccentric orbits spiralling into non-rotating
black holes {\cite{cpk}}, and was later confirmed to hold for all
spins {\cite{gk}}.)  The frequency $\Omega_\theta$ grows monotonically
over the inspiral; $\Omega_r$ reaches a maximum in the strong field,
but then turns over, approaching $0$ near the separatrix
(cf.\ Sec.\ {\ref{sec:separatrix}}).

\begin{figure}[ht]
\includegraphics[width = 0.48\textwidth]{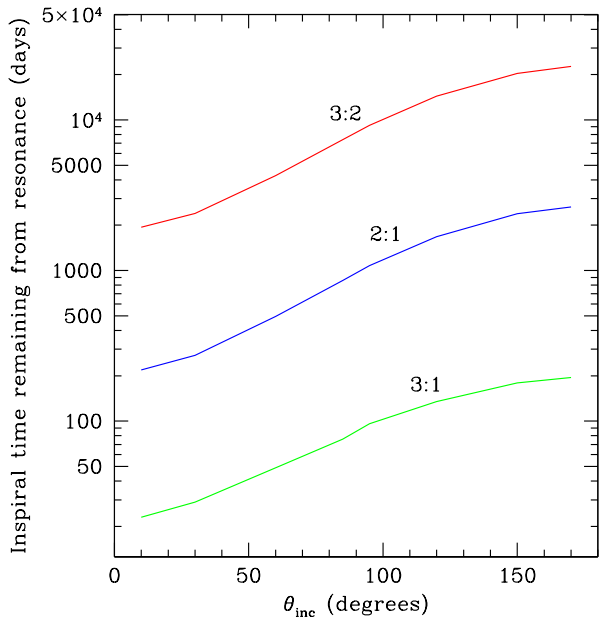}
\caption{Inspiral time remaining following resonance passage as a
  function of initial $\theta_{\rm inc}$ for $a = 0.7M$, initial $e =
  0.4$.  These numbers assume $M = 10^6\,M_\odot$, $\mu = 1\,M_\odot$;
  Using $T_{\rm insp} \propto M^2/\mu$ [Eq.\ (\ref{eq:Tinspscaling})],
  it is simple to extrapolate to other masses.  Red curve shows the
  time remaining for the 3:2 resonance, blue for the 2:1 resonance,
  and green for 3:1.  Similar results can be found for other choices
  of black hole spin and initial eccentricity; see Tables
  {\ref{tab:spacinga0.1}} -- {\ref{tab:spacinga0.9}} for detailed
  data.}
\label{fig:spacing}
\end{figure}

Figure {\ref{fig:spacing}} shows how much time remains following each
resonance for the binary shown in Fig.\ {\ref{fig:binevolve}}.  For
the masses used to generate this plot, the 3:1 resonance occurs in the
last several months of resonance at all inclinations; the 2:1
resonance occurs in the final two years of inspiral at inclinations
less than about $80^\circ$.  The 3:2 resonance occurs much earlier
(nearly six years before plunge at the largest inclination we
examine).  Bearing in mind the $M^2/\mu$ scaling of inspiral time,
this plot indicates that the 3:1 resonance occurs during the final two
years of inspiral for essentially all astrophysically important EMRIs.
In many cases, both the 2:1 and 3:1 resonances occur during this time
period.  For shallow inclination and somewhat less extreme mass ratio
(e.g., $M = 10^6\,M_\odot$, $\mu = 10\,M_\odot$), all three of the
resonances we have examined occur during the final two years of EMRI
evolution.

Data describing inspiral for other values of spin and eccentricity is
presented in Tables {\ref{tab:spacinga0.1}} --
{\ref{tab:spacinga0.9}}, in Appendix {\ref{app:timeleft}}.  Although
differing in detail, their qualitative form is broadly the same as
that shown in Fig.\ {\ref{fig:spacing}}.  In particular, for $M =
10^6\,M_\odot$, $\mu = 1\,M_\odot$, the 3:1 resonance occurs in the
last few months of inspiral nearly always, reaching to about a year
before final plunge in the most extreme cases; the 2:1 resonance
occurs during the last two years for shallow orbital inclinations and
relatively small initial eccentricites; and the 3:2 resonance occurs
several years from final plunge.  All three resonances are likely to
occur in the last few years if the smaller mass is $10\,M_\odot$ or
larger, or if the larger mass is smaller than $10^6\,M_\odot$.

\begin{figure}[ht]
\includegraphics[width = 0.48\textwidth]{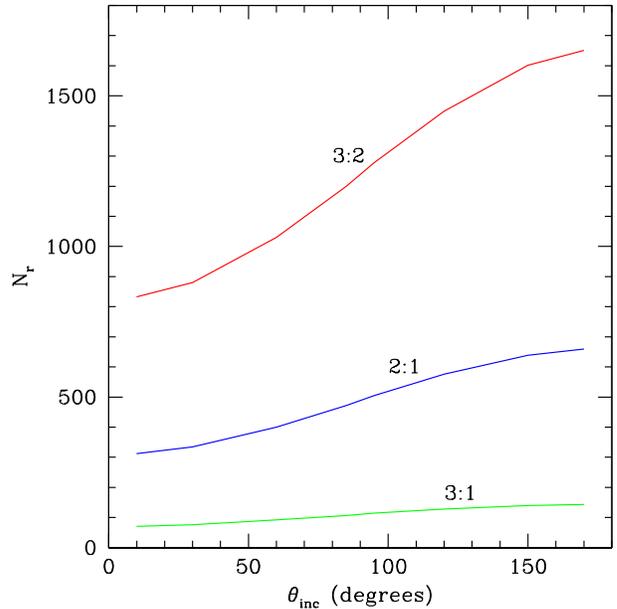}
\caption{Number of orbits $N_r$ near resonance [defined by
    Eq.\ (\ref{eq:Nthr})] as a function of initial $\theta_{\rm inc}$
  for $a = 0.7M$, initial $e = 0.4$.  These numbers assume $M =
  10^6\,M_\odot$, $\mu = 1\,M_\odot$; using the rule that $N_r \propto
  \sqrt{M/\mu}$ [Eq.\ (\ref{eq:Nthrscaling})], it is simple to
  extrapolate to other masses.  (The number of $\theta$ oscillations
  is simply related to the number of $r$ oscillations by the resonance
  number.)  The results are similar for other choices of black hole
  spin and initial orbital eccentricity.}
\label{fig:duration}
\end{figure}

Figure {\ref{fig:duration}} shows how many radial oscillations $N_r$
the system executes as it passes through each resonance, using
Eq.\ (\ref{eq:Nthr}); $N_\theta$ is simply related to $N_r$ by the
resonance number.  Computations are again done for the $a = 0.7M$,
$e_{\rm init} = 0.4$ system that was used to generate
Figs.\ {\ref{fig:statphase}}, {\ref{fig:binevolve}}, and
{\ref{fig:spacing}}.  Data describing other cases can be found in
Tables {\ref{tab:dura0.1}} -- {\ref{tab:dura0.9}} in Appendix
{\ref{app:duration}}.  In all cases, the results are qualitatively
quite similar to Fig.\ {\ref{fig:duration}}, though differing in
quantitative detail.

\begin{figure}[ht]
\includegraphics[width = 0.48\textwidth]{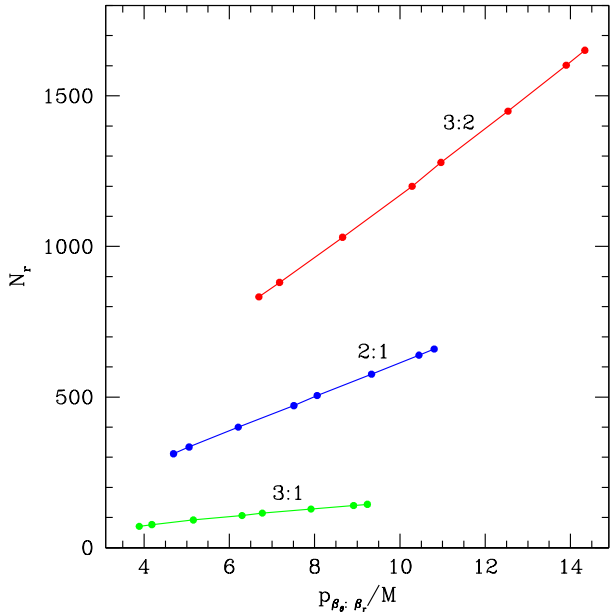}
\caption{Reorganization of the data shown in
  Fig.\ {\ref{fig:duration}}: We now show the number of orbits near
  resonance as a function of semi-latus rectum $p$ at resonance
  passage.  The points on each track label the different inclinations
  we study; from left to right, they are $\theta_{\rm inc} =
  10^\circ$, $30^\circ$, $60^\circ$, $85^\circ$, $95^\circ$,
  $120^\circ$, $150^\circ$, and $170^\circ$.  Notice the nearly linear
  dependence of $N_r$ on $p$.}
\label{fig:duration2}
\end{figure}

It's worth noting that both the duration of each resonance and the
amount of inspiral left following resonance passage depend in a very
simple way on a single parameter --- the value of semi-latus rectum
$p$ at the moment of resonance passage.  When a resonance occurs at
large $p$, EMRI systems evolve relatively slowly, and they spend more
time near resonances.  The converse holds when resonances occur at
small $p$.  We show this in Fig.\ {\ref{fig:duration2}}, which is the
same data as that shown in Fig.\ {\ref{fig:duration}}, but rearranged
to show $N_r$ as a function of $p_{\beta_\theta : \beta_r}$, the
semi-latus rectum at the moment of resonance passage.  Note that $N_r$
is close to a linear function of $p_{\beta_\theta : \beta_r}$.  This
is consistent with the results of Ref.\ {\cite{bgh}}, which shows that
eccentricity has only a modest effect on the value of $p$ at which a
resonance occurs except for orbits that are very close to the
separatrix.  A similar figure could be made to show how the time
remaining in inspiral depends on $p_{\beta_\theta : \beta_r}$.

\section{Conclusions}
\label{sec:conclude}

Taken together, Figs.\ {\ref{fig:spacing}} and {\ref{fig:duration}}
(and their counterparts in Tables {\ref{tab:spacinga0.1}} --
{\ref{tab:dura0.9}}) present the main message of this paper: Orbital
resonances are ubiquitous in extreme mass ratio inspiral, and each
resonance lasts for many oscillations --- several dozen to many
hundred, depending on the system's mass ratio.  This amplifies the
developing message of ongoing work on resonances: These effects are
likely to have an observationally important effect on astrophysical
extreme mass ratio systems.

It is worth noting that, in addition to the issues we consider here
and in associated analyses, many other factors have an impact on the
astrophysical relevance of resonances.  Among them\footnote{We thank
  this paper's referee for reminding us that the following points are
  also important, and should be emphasized.} are: The lifetime of the
eLISA mission (or whatever low-frequency GW mission eventually
measures EMRI waves); what portion of the EMRI signals overlap in time
with this mission; the strength of the waves, and hence the integrated
signal-to-noise ratio (especially the contribution to the
signal-to-noise ratio from resonances); and the EMRI event rate, both
in total, and as distributed with the masses $M$ and $\mu$.  These
points are, by and large, not well understood; and they are rather
coupled to one another.  We will not pretend to offer any deep wisdom,
but will simply note that as the community begins to think about the
impact that resonances can have upon EMRI measurements, these points
must be considered eventually as well.

Although we have now studied various pieces of what is needed to
understand the resonant evolution of extreme mass ratio binaries, it
remains necessary that they be assembled in a self-consistent manner.
Roughly speaking, the results of this paper tell us the ``width'' of
each resonance (i.e., the time in which a system's evolution is
sustantially modified by resonant physics); those of
Ref.\ {\cite{fhr}} tell us the each resonance's ``height'' (i.e., the
extent of the deviation of the on-resonance rate of change of system
parameters).  Up to a factor of order unity, the product of these
numbers tells us the kick in the binary's parameters due to evolving
through an orbital resonance (cf.\ Fig.\ 1 of Ref.\ {\cite{fh12}}).

It is satisfying that our results are, so far, consistent with the
picture originally developed by Flanagan and Hinderer {\cite{fh12}}.
However, rough agreement is not good enough for modeling EMRI systems.
More precise understanding of how a system evolves through orbital
resonances is almost certainly necessary to ensure that future
observatories like eLISA will be able to measure these waves.  Our
plan for going beyond this analysis is to compute components of the
instantaneous dissipative self force and the self consistently evolve
the system as a ``forced geodesic'' {\cite{gfdhb}}.  Results
describing the instantaneous dissipative self force for scalar fields
are given in Ref.\ {\cite{dfhcqg}}; generalizing to the gravitational
interaction is straightforward provided we focus on the dissipative
constribution.  It turns out that all of the quantities we need can be
computed with relatively minor modifications of pre-existing
Teukolsky-equation solvers {\cite{h2000,dh06}}.  As an example, one
component of the dissipative self force can be written as the rate of
change of orbital energy per unit Mino time $\lambda$
{\cite{fhhr,fhnotes}}:
\begin{equation}
\frac{dE}{d\lambda} = \sum_{k'n'} {\cal E}_{k'n'}
e^{-i(k'\Upsilon_\theta + n'\Upsilon_r)\lambda}\;,
\end{equation}
where $\Upsilon_{\theta,r}$ are orbital frequencies conjugate to
Mino-time $\lambda$, and where
\begin{equation}
{\cal E}_{k'n'} = \sum_{lmkn}\frac{\Gamma}{4\pi\omega_{mkn}^2}
Z_{lmkn}Z^*_{lm(k+k')(n+n')}\;.
\end{equation}
The amplitudes $Z_{lmkn}$ describe the amplitude of the curvature
scalar $\psi_4$ in a harmonic and Fourier decomposition; superscript
$*$ denotes complex conjugation, $\Gamma$ relates Mino-time $\lambda$
to Boyer-Lindquist time $t$, and $\omega_{mkn}$ is a mode frequency
conjugate to $t$.

The self force is of course gauge dependent; however, a two-timescale
analysis {\cite{hf08}} shows that its impact over a long timescale
$T_{\rm insp}$ is dominated by a gauge-independent result (with
subleading gauge-dependent contributions that scale as $T_{\rm
  orb}/T_{\rm insp}$).  Hence the results computed in this way are not
gauge ambiguous provided we integrate the system over a long time
compared to a typical orbital period.  We are now beginning to
overhaul our group's Teukolsky equation codebase in order to do this
analysis.

\begin{acknowledgments}
This paper originated in a question Eric Poisson asked of SAH about
whether resonances occur in realistic inspirals, or whether they are
rare curiosities; we thank him for motivating us to thoroughly examine
and answer this question.  We gratefully acknowledge many useful
discussions on the subject of resonances in extreme mass ratio
inspiral with \'Eanna Flanagan, Tanja Hinderer, and Gabriel Perez-Giz,
and we thank this paper's referee for making several helpful
suggestions.  This work was supported by NSF Grant PHY-1068720.  SAH
gratefully acknowledges fellowship support by the John Simon
Guggenheim Memorial Foundation, and sabbatical support from the
Canadian Institute for Theoretical Astrophysics and the Perimeter
Institute for Theoretical Physics.
\end{acknowledgments}

\appendix

\section{Inspiral time remaining following resonance passage}
\label{app:timeleft}

In this appendix, we present detailed data showing how the inspiral
time remaining following resonance passage depends on EMRI parameters.

\begin{table*}[ht]
\begin{ruledtabular}
\begin{tabular}{|c|c|c||c|c|c||c|c|c||c|c|c|}
\hline
$p_{\rm init}/M$ & $e_{\rm init}$ & $\theta_{\rm inc,\ init}$
& $p_{3:2}/M$ & $e_{3:2}$ & $T_{3:2}$ (days)
& $p_{2:1}/M$ & $e_{2:1}$ & $T_{2:1}$ (days)
& $p_{3:1}/M$ & $e_{3:1}$ & $T_{3:1}$ (days) \\
\hline
11.297 & 0.1 & 10$^\circ$ & 10.270 & 0.085 & 6888 & 7.580 & 0.053 & 854 & 6.390 & 0.045 & 97 \\
\hline
11.363 & 0.1 & 30$^\circ$ & 10.334 & 0.085 & 7032 & 7.630 & 0.053 & 872 & 6.433 & 0.045 & 99 \\
\hline
11.583 & 0.1 & 60$^\circ$ & 10.531 & 0.085 & 7489 & 7.786 & 0.053 & 930 & 6.566 & 0.046 & 105 \\
\hline
11.825 & 0.1 & 85$^\circ$ & 10.753 & 0.086 & 8029 & 7.961 & 0.054 & 999 & 6.718 & 0.046 & 113 \\
\hline
11.935 & 0.1 & 95$^\circ$ & 10.846 & 0.085 & 8265 & 8.035 & 0.054 & 1029 & 6.781 & 0.046 & 116 \\
\hline
12.177 & 0.1 & 120$^\circ$ & 11.067 & 0.086 & 8844 & 8.210 & 0.054 & 1103 & 6.933 & 0.047 & 125 \\
\hline
12.397 & 0.1 & 150$^\circ$ & 11.263 & 0.086 & 9379 & 8.366 & 0.054 & 1171 & 7.067 & 0.047 & 132 \\
\hline
12.463 & 0.1 & 170$^\circ$ & 11.326 & 0.086 & 9557 & 8.416 & 0.055 & 1194 & 7.111 & 0.048 & 134 \\
\hline
\hline
11.385 & 0.4 & 10$^\circ$ & 10.332 & 0.346 & 7417 & 7.624 & 0.220 & 858 & 6.460 & 0.189 & 71 \\
\hline
11.451 & 0.4 & 30$^\circ$ & 10.396 & 0.346 & 7571 & 7.674 & 0.220 & 876 & 6.504 & 0.190 & 72 \\
\hline
11.671 & 0.4 & 60$^\circ$ & 10.593 & 0.346 & 8057 & 7.831 & 0.222 & 933 & 6.640 & 0.192 & 76 \\
\hline
11.913 & 0.4 & 85$^\circ$ & 10.815 & 0.347 & 8628 & 8.007 & 0.224 & 999 & 6.795 & 0.195 & 81 \\
\hline
12.034 & 0.4 & 95$^\circ$ & 10.909 & 0.346 & 8889 & 8.081 & 0.223 & 1031 & 6.858 & 0.194 & 84 \\
\hline
12.276 & 0.4 & 120$^\circ$ & 11.130 & 0.346 & 9500 & 8.258 & 0.225 & 1101 & 7.013 & 0.197 & 88 \\
\hline
12.485 & 0.4 & 150$^\circ$ & 11.326 & 0.347 & 10072 & 8.414 & 0.227 & 1167 & 7.151 & 0.200 & 93 \\
\hline
12.562 & 0.4 & 170$^\circ$ & 11.389 & 0.346 & 10260 & 8.465 & 0.227 & 1189 & 7.195 & 0.200 & 94 \\
\hline
\hline
11.572 & 0.7 & 10$^\circ$ & 10.473 & 0.618 & 9588 & 7.742 & 0.419 & 929 & 6.648 & 0.370 & 49 \\
\hline
11.649 & 0.7 & 30$^\circ$ & 10.537 & 0.618 & 9775 & 7.794 & 0.420 & 947 & 6.693 & 0.371 & 49 \\
\hline
11.869 & 0.7 & 60$^\circ$ & 10.734 & 0.618 & 10390 & 7.952 & 0.422 & 1006 & 6.834 & 0.375 & 52 \\
\hline
12.111 & 0.7 & 85$^\circ$ & 10.957 & 0.618 & 11108 & 8.131 & 0.426 & 1071 & 6.996 & 0.380 & 53 \\
\hline
12.221 & 0.7 & 95$^\circ$ & 11.051 & 0.618 & 11460 & 8.205 & 0.424 & 1110 & 7.059 & 0.378 & 56 \\
\hline
12.474 & 0.7 & 120$^\circ$ & 11.273 & 0.618 & 12218 & 8.384 & 0.427 & 1179 & 7.221 & 0.383 & 58 \\
\hline
12.683 & 0.7 & 150$^\circ$ & 11.470 & 0.618 & 12956 & 8.543 & 0.430 & 1248 & 7.365 & 0.387 & 61 \\
\hline
12.760 & 0.7 & 170$^\circ$ & 11.533 & 0.618 & 13189 & 8.594 & 0.430 & 1270 & 7.411 & 0.388 & 62 \\
\hline
\hline
11.803 & 0.95 & 10$^\circ$ & 10.651 & 0.853 & 18801 & 7.926 & 0.622 & 1139 & 6.933 & 0.569 & 35 \\
\hline
11.880 & 0.95 & 30$^\circ$ & 10.715 & 0.852 & 19128 & 7.978 & 0.622 & 1160 & 6.979 & 0.570 & 36 \\
\hline
12.100 & 0.95 & 60$^\circ$ & 10.913 & 0.852 & 20320 & 8.138 & 0.624 & 1230 & 7.125 & 0.574 & 37 \\
\hline
12.342 & 0.95 & 85$^\circ$ & 11.137 & 0.853 & 21773 & 8.320 & 0.629 & 1308 & 7.294 & 0.580 & 39 \\
\hline
12.452 & 0.95 & 95$^\circ$ & 11.230 & 0.852 & 22396 & 8.393 & 0.626 & 1359 & 7.356 & 0.577 & 41 \\
\hline
12.705 & 0.95 & 120$^\circ$ & 11.453 & 0.852 & 23851 & 8.575 & 0.630 & 1441 & 7.524 & 0.582 & 42 \\
\hline
12.925 & 0.95 & 150$^\circ$ & 11.651 & 0.852 & 25264 & 8.736 & 0.632 & 1524 & 7.672 & 0.586 & 44 \\
\hline
12.991 & 0.95 & 170$^\circ$ & 11.715 & 0.852 & 25793 & 8.789 & 0.633 & 1552 & 7.720 & 0.587 & 45 \\
\hline
\end{tabular}
\end{ruledtabular}
\caption{Various chararacteristics of inspirals that encounter the
  3:2, 2:1, and 3:1 orbital resonances.  We show the values of $p$,
  $e$, and $\theta_{\rm inc}$ at the beginning of each inspiral, and
  the values of $p$, $e$, and the time remaining in the inspiral at
  each resonance crossing.  (The inclination $\theta_{\rm inc}$ barely
  changes during inspiral, so we do not show those data.)  In all
  cases, the large black hole's spin is $a = 0.1M$; the binary's
  masses are $M = 10^6\,M_\odot$, $\mu = 1\,M_\odot$.  The times can
  be extrapolated to other masses using the rule that $T_{\rm insp}
  \propto M^2/\mu$ [Eq.\ (\ref{eq:Tinspscaling})].  Notice that
  significant inspiral remains following each resonance.}
\label{tab:spacinga0.1}
\end{table*}

\begin{table*}[ht]
\begin{ruledtabular}
\begin{tabular}{|c|c|c||c|c|c||c|c|c||c|c|c|}
\hline
$p_{\rm init}/M$ & $e_{\rm init}$ & $\theta_{\rm inc,\ init}$
& $p_{3:2}/M$ & $e_{3:2}$ & $T_{3:2}$ (days)
& $p_{2:1}/M$ & $e_{2:1}$ & $T_{2:1}$ (days)
& $p_{3:1}/M$ & $e_{3:1}$ & $T_{3:1}$ (days) \\
\hline
9.427 & 0.1 & 10$^\circ$ & 8.574 & 0.084 & 3844 & 6.229 & 0.048 & 468 & 5.223 & 0.038 & 54 \\
\hline
9.713 & 0.1 & 30$^\circ$ & 8.836 & 0.085 & 4240 & 6.432 & 0.048 & 518 & 5.393 & 0.039 & 59 \\
\hline
10.604 & 0.1 & 60$^\circ$ & 9.642 & 0.085 & 5642 & 7.061 & 0.050 & 694 & 5.926 & 0.042 & 79 \\
\hline
11.594 & 0.1 & 85$^\circ$ & 10.542 & 0.085 & 7575 & 7.772 & 0.053 & 939 & 6.538 & 0.045 & 106 \\
\hline
12.012 & 0.1 & 95$^\circ$ & 10.919 & 0.085 & 8519 & 8.072 & 0.053 & 1060 & 6.797 & 0.045 & 120 \\
\hline
12.991 & 0.1 & 120$^\circ$ & 11.805 & 0.086 & 11063 & 8.781 & 0.055 & 1386 & 7.414 & 0.048 & 156 \\
\hline
13.849 & 0.1 & 150$^\circ$ & 12.582 & 0.086 & 13721 & 9.406 & 0.056 & 1729 & 7.961 & 0.050 & 194 \\
\hline
14.124 & 0.1 & 170$^\circ$ & 12.832 & 0.086 & 14670 & 9.608 & 0.057 & 1851 & 8.138 & 0.051 & 207 \\
\hline
\hline
9.515 & 0.4 & 10$^\circ$ & 8.632 & 0.343 & 4166 & 6.263 & 0.200 & 482 & 5.269 & 0.161 & 46 \\
\hline
9.801 & 0.4 & 30$^\circ$ & 8.894 & 0.343 & 4584 & 6.467 & 0.203 & 529 & 5.441 & 0.165 & 48 \\
\hline
10.681 & 0.4 & 60$^\circ$ & 9.699 & 0.345 & 6071 & 7.099 & 0.212 & 703 & 5.983 & 0.177 & 61 \\
\hline
11.671 & 0.4 & 85$^\circ$ & 10.601 & 0.347 & 8108 & 7.816 & 0.221 & 938 & 6.608 & 0.191 & 77 \\
\hline
12.100 & 0.4 & 95$^\circ$ & 10.978 & 0.346 & 9157 & 8.116 & 0.221 & 1067 & 6.868 & 0.189 & 90 \\
\hline
13.079 & 0.4 & 120$^\circ$ & 11.867 & 0.347 & 11851 & 8.830 & 0.229 & 1378 & 7.500 & 0.202 & 110 \\
\hline
13.937 & 0.4 & 150$^\circ$ & 12.647 & 0.348 & 14688 & 9.459 & 0.234 & 1709 & 8.059 & 0.210 & 132 \\
\hline
14.223 & 0.4 & 170$^\circ$ & 12.898 & 0.348 & 15699 & 9.663 & 0.235 & 1826 & 8.241 & 0.212 & 139 \\
\hline
\hline
9.702 & 0.7 & 10$^\circ$ & 8.764 & 0.614 & 5419 & 6.360 & 0.388 & 536 & 5.400 & 0.321 & 34 \\
\hline
9.977 & 0.7 & 30$^\circ$ & 9.026 & 0.616 & 5954 & 6.566 & 0.393 & 589 & 5.580 & 0.329 & 37 \\
\hline
10.868 & 0.7 & 60$^\circ$ & 9.830 & 0.617 & 7812 & 7.205 & 0.406 & 768 & 6.142 & 0.349 & 45 \\
\hline
11.858 & 0.7 & 85$^\circ$ & 10.734 & 0.619 & 10355 & 7.932 & 0.423 & 1002 & 6.796 & 0.375 & 53 \\
\hline
12.287 & 0.7 & 95$^\circ$ & 11.112 & 0.617 & 11767 & 8.232 & 0.419 & 1155 & 7.055 & 0.369 & 62 \\
\hline
13.266 & 0.7 & 120$^\circ$ & 12.007 & 0.619 & 15178 & 8.959 & 0.433 & 1468 & 7.719 & 0.391 & 72 \\
\hline
14.135 & 0.7 & 150$^\circ$ & 12.792 & 0.620 & 18807 & 9.599 & 0.441 & 1804 & 8.303 & 0.403 & 83 \\
\hline
14.421 & 0.7 & 170$^\circ$ & 13.046 & 0.620 & 20100 & 9.806 & 0.443 & 1925 & 8.493 & 0.407 & 87 \\
\hline
\hline
9.911 & 0.95 & 10$^\circ$ & 8.935 & 0.851 & 10636 & 6.523 & 0.591 & 671 & 5.636 & 0.516 & 26 \\
\hline
10.197 & 0.95 & 30$^\circ$ & 9.195 & 0.851 & 11629 & 6.730 & 0.596 & 730 & 5.820 & 0.524 & 28 \\
\hline
11.077 & 0.95 & 60$^\circ$ & 9.998 & 0.852 & 15228 & 7.374 & 0.609 & 940 & 6.400 & 0.547 & 33 \\
\hline
12.078 & 0.95 & 85$^\circ$ & 10.903 & 0.854 & 20178 & 8.111 & 0.627 & 1208 & 7.081 & 0.576 & 38 \\
\hline
12.507 & 0.95 & 95$^\circ$ & 11.283 & 0.851 & 22794 & 8.410 & 0.620 & 1416 & 7.336 & 0.564 & 47 \\
\hline
13.497 & 0.95 & 120$^\circ$ & 12.183 & 0.853 & 29589 & 9.150 & 0.635 & 1789 & 8.028 & 0.589 & 52 \\
\hline
14.377 & 0.95 & 150$^\circ$ & 12.975 & 0.854 & 36785 & 9.802 & 0.644 & 2203 & 8.636 & 0.602 & 60 \\
\hline
14.663 & 0.95 & 170$^\circ$ & 13.231 & 0.854 & 39357 & 10.014 & 0.646 & 2352 & 8.834 & 0.606 & 62 \\
\hline
\end{tabular}
\end{ruledtabular}
\caption{Same as Table {\ref{tab:spacinga0.1}}, but for
  $a = 0.4M$.}
\label{tab:spacinga0.4}
\end{table*}

\begin{table*}[ht]
\begin{ruledtabular}
\begin{tabular}{|c|c|c||c|c|c||c|c|c||c|c|c|}
\hline
$p_{\rm init}/M$ & $e_{\rm init}$ & $\theta_{\rm inc,\ init}$
& $p_{3:2}/M$ & $e_{3:2}$ & $T_{3:2}$ (days)
& $p_{2:1}/M$ & $e_{2:1}$ & $T_{2:1}$ (days)
& $p_{3:1}/M$ & $e_{3:1}$ & $T_{3:1}$ (days) \\
\hline
7.304 & 0.1 & 10$^\circ$ & 6.642 & 0.082 & 1771 & 4.670 & 0.037 & 210 & 3.870 & 0.025 & 25 \\
\hline
7.832 & 0.1 & 30$^\circ$ & 7.127 & 0.083 & 2203 & 5.033 & 0.039 & 262 & 4.162 & 0.027 & 30 \\
\hline
9.460 & 0.1 & 60$^\circ$ & 8.607 & 0.084 & 3984 & 6.180 & 0.045 & 483 & 5.120 & 0.034 & 57 \\
\hline
11.253 & 0.1 & 85$^\circ$ & 10.236 & 0.085 & 6964 & 7.477 & 0.051 & 856 & 6.239 & 0.042 & 97 \\
\hline
12.001 & 0.1 & 95$^\circ$ & 10.911 & 0.085 & 8596 & 8.019 & 0.051 & 1064 & 6.713 & 0.043 & 121 \\
\hline
13.728 & 0.1 & 120$^\circ$ & 12.478 & 0.086 & 13452 & 9.286 & 0.055 & 1688 & 7.827 & 0.048 & 190 \\
\hline
15.224 & 0.1 & 150$^\circ$ & 13.836 & 0.086 & 19094 & 10.390 & 0.058 & 2422 & 8.803 & 0.052 & 270 \\
\hline
15.708 & 0.1 & 170$^\circ$ & 14.271 & 0.086 & 21226 & 10.744 & 0.058 & 2701 & 9.117 & 0.053 & 301 \\
\hline
\hline
7.381 & 0.4 & 10$^\circ$ & 6.693 & 0.336 & 1936 & 4.689 & 0.158 & 219 & 3.887 & 0.104 & 23 \\
\hline
7.909 & 0.4 & 30$^\circ$ & 7.175 & 0.338 & 2390 & 5.055 & 0.167 & 273 & 4.183 & 0.114 & 29 \\
\hline
9.526 & 0.4 & 60$^\circ$ & 8.654 & 0.343 & 4268 & 6.209 & 0.192 & 495 & 5.155 & 0.146 & 49 \\
\hline
11.319 & 0.4 & 85$^\circ$ & 10.286 & 0.346 & 7401 & 7.514 & 0.215 & 859 & 6.296 & 0.180 & 76 \\
\hline
12.078 & 0.4 & 95$^\circ$ & 10.962 & 0.345 & 9209 & 8.058 & 0.215 & 1078 & 6.773 & 0.180 & 96 \\
\hline
13.816 & 0.4 & 120$^\circ$ & 12.536 & 0.347 & 14356 & 9.335 & 0.230 & 1677 & 7.914 & 0.203 & 135 \\
\hline
15.323 & 0.4 & 150$^\circ$ & 13.900 & 0.348 & 20383 & 10.447 & 0.239 & 2379 & 8.914 & 0.217 & 179 \\
\hline
15.807 & 0.4 & 170$^\circ$ & 14.338 & 0.349 & 22661 & 10.804 & 0.241 & 2644 & 9.236 & 0.220 & 195 \\
\hline
\hline
7.546 & 0.7 & 10$^\circ$ & 6.811 & 0.608 & 2564 & 4.749 & 0.315 & 258 & 3.938 & 0.210 & 24 \\
\hline
8.063 & 0.7 & 30$^\circ$ & 7.290 & 0.611 & 3118 & 5.122 & 0.333 & 316 & 4.247 & 0.232 & 28 \\
\hline
9.669 & 0.7 & 60$^\circ$ & 8.764 & 0.616 & 5442 & 6.290 & 0.377 & 552 & 5.259 & 0.297 & 42 \\
\hline
11.473 & 0.7 & 85$^\circ$ & 10.401 & 0.620 & 9311 & 7.614 & 0.415 & 912 & 6.454 & 0.359 & 54 \\
\hline
12.232 & 0.7 & 95$^\circ$ & 11.080 & 0.617 & 11742 & 8.160 & 0.411 & 1175 & 6.933 & 0.353 & 72 \\
\hline
13.992 & 0.7 & 120$^\circ$ & 12.666 & 0.620 & 18249 & 9.460 & 0.435 & 1778 & 8.132 & 0.393 & 90 \\
\hline
15.521 & 0.7 & 150$^\circ$ & 14.046 & 0.621 & 25983 & 10.594 & 0.448 & 2493 & 9.181 & 0.414 & 111 \\
\hline
16.005 & 0.7 & 170$^\circ$ & 14.488 & 0.621 & 28936 & 10.959 & 0.452 & 2762 & 9.520 & 0.420 & 118 \\
\hline
\hline
7.744 & 0.95 & 10$^\circ$ & 6.967 & 0.850 & 5115 & 4.874 & 0.515 & 365 & 4.052 & 0.354 & 30 \\
\hline
8.250 & 0.95 & 30$^\circ$ & 7.440 & 0.851 & 6120 & 5.250 & 0.534 & 428 & 4.387 & 0.394 & 33 \\
\hline
9.845 & 0.95 & 60$^\circ$ & 8.905 & 0.854 & 10475 & 6.426 & 0.579 & 681 & 5.451 & 0.485 & 35 \\
\hline
11.660 & 0.95 & 85$^\circ$ & 10.546 & 0.856 & 17951 & 7.768 & 0.620 & 1082 & 6.703 & 0.562 & 39 \\
\hline
12.419 & 0.95 & 95$^\circ$ & 11.229 & 0.853 & 22546 & 8.315 & 0.610 & 1437 & 7.181 & 0.544 & 56 \\
\hline
14.201 & 0.95 & 120$^\circ$ & 12.831 & 0.854 & 35426 & 9.643 & 0.637 & 2159 & 8.438 & 0.591 & 66 \\
\hline
15.752 & 0.95 & 150$^\circ$ & 14.228 & 0.855 & 50934 & 10.804 & 0.652 & 3045 & 9.535 & 0.614 & 80 \\
\hline
16.247 & 0.95 & 170$^\circ$ & 14.677 & 0.855 & 56940 & 11.179 & 0.656 & 3384 & 9.890 & 0.620 & 85 \\
\hline
\end{tabular}
\end{ruledtabular}
\caption{Same as Table {\ref{tab:spacinga0.1}}, but for
  $a = 0.7M$.}
\label{tab:spacinga0.7}
\end{table*}

\begin{table*}[ht]
\begin{ruledtabular}
\begin{tabular}{|c|c|c||c|c|c||c|c|c||c|c|c|}
\hline
$p_{\rm init}/M$ & $e_{\rm init}$ & $\theta_{\rm inc,\ init}$
& $p_{3:2}/M$ & $e_{3:2}$ & $T_{3:2}$ (days)
& $p_{2:1}/M$ & $e_{2:1}$ & $T_{2:1}$ (days)
& $p_{3:1}/M$ & $e_{3:1}$ & $T_{3:1}$ (days) \\
\hline
5.555 & 0.1 & 10$^\circ$ & 5.053 & 0.079 & 867 & 3.323 & 0.019 & 103 & 2.680 & 0.007 & 13 \\
\hline
6.303 & 0.1 & 30$^\circ$ & 5.737 & 0.080 & 1228 & 3.825 & 0.025 & 144 & 3.057 & 0.010 & 17 \\
\hline
8.547 & 0.1 & 60$^\circ$ & 7.788 & 0.083 & 2992 & 5.438 & 0.039 & 356 & 4.403 & 0.025 & 42 \\
\hline
10.956 & 0.1 & 85$^\circ$ & 9.971 & 0.085 & 6481 & 7.206 & 0.049 & 790 & 5.953 & 0.039 & 91 \\
\hline
11.946 & 0.1 & 95$^\circ$ & 10.858 & 0.085 & 8555 & 7.929 & 0.050 & 1054 & 6.593 & 0.040 & 121 \\
\hline
14.190 & 0.1 & 120$^\circ$ & 12.896 & 0.086 & 15142 & 9.592 & 0.055 & 1901 & 8.070 & 0.048 & 214 \\
\hline
16.115 & 0.1 & 150$^\circ$ & 14.644 & 0.086 & 23311 & 11.021 & 0.058 & 2968 & 9.343 & 0.053 & 331 \\
\hline
16.731 & 0.1 & 170$^\circ$ & 15.201 & 0.086 & 26496 & 11.478 & 0.059 & 3386 & 9.749 & 0.054 & 376 \\
\hline
\hline
5.621 & 0.4 & 10$^\circ$ & 5.092 & 0.323 & 950 & 3.329 & 0.081 & 106 & 2.682 & 0.027 & 14 \\
\hline
6.358 & 0.4 & 30$^\circ$ & 5.775 & 0.330 & 1326 & 3.838 & 0.107 & 149 & 3.065 & 0.044 & 18 \\
\hline
8.602 & 0.4 & 60$^\circ$ & 7.824 & 0.340 & 3180 & 5.460 & 0.167 & 369 & 4.424 & 0.107 & 41 \\
\hline
11.011 & 0.4 & 85$^\circ$ & 10.012 & 0.346 & 6838 & 7.237 & 0.207 & 797 & 5.997 & 0.167 & 75 \\
\hline
12.001 & 0.4 & 95$^\circ$ & 10.902 & 0.345 & 9127 & 7.963 & 0.210 & 1072 & 6.643 & 0.170 & 100 \\
\hline
14.267 & 0.4 & 120$^\circ$ & 12.950 & 0.347 & 16117 & 9.639 & 0.230 & 1889 & 8.155 & 0.202 & 154 \\
\hline
16.203 & 0.4 & 150$^\circ$ & 14.708 & 0.349 & 24847 & 11.081 & 0.242 & 2906 & 9.461 & 0.220 & 217 \\
\hline
16.830 & 0.4 & 170$^\circ$ & 15.269 & 0.349 & 28253 & 11.541 & 0.244 & 3301 & 9.879 & 0.225 & 239 \\
\hline
\hline
5.753 & 0.7 & 10$^\circ$ & 5.190 & 0.597 & 1266 & 3.346 & 0.164 & 114 & 2.687 & 0.053 & 13 \\
\hline
6.479 & 0.7 & 30$^\circ$ & 5.866 & 0.605 & 1709 & 3.876 & 0.222 & 166 & 3.086 & 0.090 & 19 \\
\hline
8.701 & 0.7 & 60$^\circ$ & 7.908 & 0.616 & 3976 & 5.521 & 0.337 & 414 & 4.484 & 0.223 & 41 \\
\hline
11.132 & 0.7 & 85$^\circ$ & 10.105 & 0.620 & 8469 & 7.320 & 0.404 & 845 & 6.124 & 0.339 & 57 \\
\hline
12.133 & 0.7 & 95$^\circ$ & 11.002 & 0.617 & 11526 & 8.051 & 0.402 & 1172 & 6.779 & 0.336 & 81 \\
\hline
14.421 & 0.7 & 120$^\circ$ & 13.072 & 0.621 & 20388 & 9.759 & 0.436 & 1998 & 8.369 & 0.392 & 103 \\
\hline
16.401 & 0.7 & 150$^\circ$ & 14.852 & 0.622 & 31606 & 11.231 & 0.453 & 3032 & 9.742 & 0.420 & 132 \\
\hline
17.028 & 0.7 & 170$^\circ$ & 15.420 & 0.622 & 36025 & 11.702 & 0.457 & 3435 & 10.182 & 0.427 & 143 \\
\hline
\hline
5.918 & 0.95 & 10$^\circ$ & 5.331 & 0.856 & 2719 & 3.391 & 0.292 & 140 & 2.697 & 0.084 & 15 \\
\hline
6.622 & 0.95 & 30$^\circ$ & 5.991 & 0.855 & 3399 & 3.959 & 0.395 & 218 & 3.127 & 0.152 & 22 \\
\hline
8.833 & 0.95 & 60$^\circ$ & 8.016 & 0.856 & 7520 & 5.622 & 0.535 & 524 & 4.601 & 0.380 & 47 \\
\hline
11.275 & 0.95 & 85$^\circ$ & 10.224 & 0.858 & 16165 & 7.448 & 0.611 & 987 & 6.332 & 0.541 & 43 \\
\hline
12.287 & 0.95 & 95$^\circ$ & 11.129 & 0.854 & 21898 & 8.184 & 0.600 & 1427 & 6.992 & 0.523 & 66 \\
\hline
14.619 & 0.95 & 120$^\circ$ & 13.225 & 0.855 & 39459 & 9.933 & 0.638 & 2417 & 8.666 & 0.590 & 76 \\
\hline
16.632 & 0.95 & 150$^\circ$ & 15.032 & 0.856 & 62016 & 11.445 & 0.657 & 3706 & 10.107 & 0.620 & 96 \\
\hline
17.270 & 0.95 & 170$^\circ$ & 15.610 & 0.856 & 71082 & 11.929 & 0.662 & 4214 & 10.570 & 0.628 & 104 \\
\hline
\end{tabular}
\end{ruledtabular}
\caption{Same as Table {\ref{tab:spacinga0.1}}, but for
  $a = 0.9M$.}
\label{tab:spacinga0.9}
\end{table*}

\section{Detailed tables describing duration of each resonance}
\label{app:duration}

In this appendix, we present detailed data showing how the duration of
each resonance depends on EMRI parameters.

\begin{table*}[ht]
\begin{ruledtabular}
\begin{tabular}{|c|c|c|c|c|c|c|c|c|}
\hline
$p_{\rm init}/M$ & $e_{\rm init}$ & $\theta_{\rm inc,\ init}$
& $N_{\theta,3:2}$ & $N_{r,3:2}$ & $N_{\theta,2:1}$ & $N_{r,2:1}$ 
& $N_{\theta,3:1}$ & $N_{r,3:1}$ \\
\hline
11.297 & 0.1 & 10$^\circ$ & 1827.94 & 1218.62 & 930.84 & 465.42 & 323.97 & 107.99 \\
\hline
11.363 & 0.1 & 30$^\circ$ & 1838.92 & 1225.95 & 936.91 & 468.45 & 326.14 & 108.71 \\
\hline
11.583 & 0.1 & 60$^\circ$ & 1872.76 & 1248.51 & 955.57 & 477.79 & 332.78 & 110.93 \\
\hline
11.825 & 0.1 & 85$^\circ$ & 1910.84 & 1273.89 & 976.54 & 488.27 & 340.22 & 113.41 \\
\hline
11.935 & 0.1 & 95$^\circ$ & 1927.01 & 1284.68 & 985.45 & 492.72 & 343.41 & 114.47 \\
\hline
12.177 & 0.1 & 120$^\circ$ & 1965.02 & 1310.01 & 1006.34 & 503.17 & 350.80 & 116.93 \\
\hline
12.397 & 0.1 & 150$^\circ$ & 1998.72 & 1332.48 & 1024.86 & 512.43 & 357.35 & 119.12 \\
\hline
12.463 & 0.1 & 170$^\circ$ & 2009.64 & 1339.76 & 1030.86 & 515.43 & 359.47 & 119.82 \\
\hline
\hline
11.385 & 0.4 & 10$^\circ$ & 1795.66 & 1197.10 & 936.83 & 468.42 & 312.82 & 104.27 \\
\hline
11.451 & 0.4 & 30$^\circ$ & 1806.08 & 1204.05 & 942.73 & 471.37 & 314.64 & 104.88 \\
\hline
11.671 & 0.4 & 60$^\circ$ & 1838.33 & 1225.56 & 960.83 & 480.42 & 320.26 & 106.75 \\
\hline
11.913 & 0.4 & 85$^\circ$ & 1874.19 & 1249.46 & 980.84 & 490.42 & 326.12 & 108.71 \\
\hline
12.034 & 0.4 & 95$^\circ$ & 1891.08 & 1260.72 & 990.38 & 495.19 & 329.75 & 109.92 \\
\hline
12.276 & 0.4 & 120$^\circ$ & 1926.99 & 1284.66 & 1010.35 & 505.17 & 335.53 & 111.84 \\
\hline
12.485 & 0.4 & 150$^\circ$ & 1959.24 & 1306.16 & 1028.34 & 514.17 & 340.88 & 113.63 \\
\hline
12.562 & 0.4 & 170$^\circ$ & 1969.80 & 1313.20 & 1034.17 & 517.09 & 342.64 & 114.21 \\
\hline
\hline
11.572 & 0.7 & 10$^\circ$ & 1742.62 & 1161.75 & 946.20 & 473.10 & 284.55 &  94.85 \\
\hline
11.649 & 0.7 & 30$^\circ$ & 1752.20 & 1168.13 & 951.58 & 475.79 & 285.79 &  95.26 \\
\hline
11.869 & 0.7 & 60$^\circ$ & 1781.53 & 1187.69 & 968.11 & 484.06 & 289.31 &  96.44 \\
\hline
12.111 & 0.7 & 85$^\circ$ & 1813.42 & 1208.95 & 985.49 & 492.74 & 291.96 &  97.32 \\
\hline
12.221 & 0.7 & 95$^\circ$ & 1830.97 & 1220.65 & 996.76 & 498.38 & 296.72 &  98.91 \\
\hline
12.474 & 0.7 & 120$^\circ$ & 1863.26 & 1242.17 & 1014.20 & 507.10 & 299.42 &  99.81 \\
\hline
12.683 & 0.7 & 150$^\circ$ & 1892.85 & 1261.90 & 1030.74 & 515.37 & 302.55 & 100.85 \\
\hline
12.760 & 0.7 & 170$^\circ$ & 1902.64 & 1268.42 & 1036.13 & 518.07 & 303.68 & 101.23 \\
\hline
\hline
11.803 & 0.95 & 10$^\circ$ & 1707.47 & 1138.31 & 945.63 & 472.81 & 242.47 &  80.82 \\
\hline
11.880 & 0.95 & 30$^\circ$ & 1716.08 & 1144.05 & 950.38 & 475.19 & 243.25 &  81.08 \\
\hline
12.100 & 0.95 & 60$^\circ$ & 1742.52 & 1161.68 & 964.87 & 482.43 & 245.22 &  81.74 \\
\hline
12.342 & 0.95 & 85$^\circ$ & 1770.70 & 1180.47 & 978.79 & 489.40 & 245.45 &  81.82 \\
\hline
12.452 & 0.95 & 95$^\circ$ & 1788.46 & 1192.30 & 992.34 & 496.17 & 251.34 &  83.78 \\
\hline
12.705 & 0.95 & 120$^\circ$ & 1817.11 & 1211.40 & 1006.51 & 503.26 & 251.75 &  83.92 \\
\hline
12.925 & 0.95 & 150$^\circ$ & 1844.22 & 1229.48 & 1021.19 & 510.59 & 253.56 &  84.52 \\
\hline
12.991 & 0.95 & 170$^\circ$ & 1853.01 & 1235.34 & 1025.96 & 512.98 & 254.08 &  84.69 \\
\hline

\end{tabular}
\end{ruledtabular}
\caption{Number of $\theta$ and $r$ oscillations spent near resonance
  as a function of initial orbit parameters for spin $a = 0.1M$,
  defined by Eq.\ (\ref{eq:Nthr}).  All values are for $M =
  10^6\,M_\odot$, $\mu = 1\,M_\odot$; these results can be
  extrapolated to other masses using the rule that $N_{\theta,r}
  \propto \sqrt{M/\mu}$ [Eq.\ (\ref{eq:Nthrscaling})].  Notice that
  the number of oscillations near each resonance is quite large:
  $\sim100$ in the shortest cases, and $\sim 2000$ in the longest
  ones.}
\label{tab:dura0.1}
\end{table*}

\begin{table*}[ht]
\begin{ruledtabular}
\begin{tabular}{|c|c|c|c|c|c|c|c|c|}
\hline
$p_{\rm init}/M$ & $e_{\rm init}$ & $\theta_{\rm inc,\ init}$
& $N_{\theta,3:2}$ & $N_{r,3:2}$ & $N_{\theta,2:1}$ & $N_{r,2:1}$ 
& $N_{\theta,3:1}$ & $N_{r,3:1}$ \\
\hline
9.427 & 0.1 & 10$^\circ$ & 1548.67 & 1032.45 & 776.52 & 388.26 & 268.49 &  89.50 \\
\hline
9.713 & 0.1 & 30$^\circ$ & 1593.30 & 1062.20 & 801.38 & 400.69 & 277.47 &  92.49 \\
\hline
10.604 & 0.1 & 60$^\circ$ & 1729.99 & 1153.33 & 877.14 & 438.57 & 304.71 & 101.57 \\
\hline
11.594 & 0.1 & 85$^\circ$ & 1882.99 & 1255.32 & 961.58 & 480.79 & 334.85 & 111.62 \\
\hline
12.012 & 0.1 & 95$^\circ$ & 1947.76 & 1298.51 & 997.32 & 498.66 & 347.73 & 115.91 \\
\hline
12.991 & 0.1 & 120$^\circ$ & 2099.66 & 1399.78 & 1081.05 & 540.53 & 377.50 & 125.83 \\
\hline
13.849 & 0.1 & 150$^\circ$ & 2233.88 & 1489.25 & 1155.01 & 577.51 & 403.78 & 134.59 \\
\hline
14.124 & 0.1 & 170$^\circ$ & 2277.32 & 1518.21 & 1178.95 & 589.47 & 412.27 & 137.42 \\
\hline
\hline
9.515 & 0.4 & 10$^\circ$ & 1529.13 & 1019.42 & 786.66 & 393.33 & 266.04 &  88.68 \\
\hline
9.801 & 0.4 & 30$^\circ$ & 1570.73 & 1047.15 & 810.73 & 405.36 & 274.16 &  91.39 \\
\hline
10.681 & 0.4 & 60$^\circ$ & 1699.28 & 1132.85 & 884.09 & 442.05 & 298.16 &  99.39 \\
\hline
11.671 & 0.4 & 85$^\circ$ & 1842.97 & 1228.65 & 964.62 & 482.31 & 322.46 & 107.49 \\
\hline
12.100 & 0.4 & 95$^\circ$ & 1910.21 & 1273.47 & 1002.91 & 501.45 & 336.62 & 112.21 \\
\hline
13.079 & 0.4 & 120$^\circ$ & 2054.55 & 1369.70 & 1083.13 & 541.56 & 359.76 & 119.92 \\
\hline
13.937 & 0.4 & 150$^\circ$ & 2184.31 & 1456.20 & 1155.10 & 577.55 & 380.60 & 126.87 \\
\hline
14.223 & 0.4 & 170$^\circ$ & 2226.53 & 1484.36 & 1178.41 & 589.21 & 387.30 & 129.10 \\
\hline
\hline
9.702 & 0.7 & 10$^\circ$ & 1499.32 & 999.55 & 809.12 & 404.56 & 258.25 &  86.08 \\
\hline
9.977 & 0.7 & 30$^\circ$ & 1535.56 & 1023.70 & 830.71 & 415.35 & 263.82 &  87.94 \\
\hline
10.868 & 0.7 & 60$^\circ$ & 1649.91 & 1099.94 & 896.71 & 448.36 & 279.47 &  93.16 \\
\hline
11.858 & 0.7 & 85$^\circ$ & 1777.12 & 1184.74 & 966.22 & 483.11 & 290.31 &  96.77 \\
\hline
12.287 & 0.7 & 95$^\circ$ & 1847.39 & 1231.59 & 1011.20 & 505.60 & 308.72 & 102.91 \\
\hline
13.266 & 0.7 & 120$^\circ$ & 1977.83 & 1318.56 & 1081.91 & 540.96 & 318.66 & 106.22 \\
\hline
14.135 & 0.7 & 150$^\circ$ & 2098.99 & 1399.33 & 1148.54 & 574.27 & 330.38 & 110.13 \\
\hline
14.421 & 0.7 & 170$^\circ$ & 2138.64 & 1425.76 & 1170.24 & 585.12 & 334.21 & 111.40 \\
\hline
\hline
9.911 & 0.95 & 10$^\circ$ & 1486.71 & 991.14 & 827.67 & 413.84 & 235.40 &  78.47 \\
\hline
10.197 & 0.95 & 30$^\circ$ & 1518.30 & 1012.20 & 845.36 & 422.68 & 237.80 &  79.27 \\
\hline
11.077 & 0.95 & 60$^\circ$ & 1619.26 & 1079.51 & 901.09 & 450.55 & 244.46 &  81.49 \\
\hline
12.078 & 0.95 & 85$^\circ$ & 1731.22 & 1154.15 & 955.86 & 477.93 & 243.91 &  81.30 \\
\hline
12.507 & 0.95 & 95$^\circ$ & 1802.38 & 1201.58 & 1009.73 & 504.87 & 267.11 &  89.04 \\
\hline
13.497 & 0.95 & 120$^\circ$ & 1919.46 & 1279.64 & 1067.84 & 533.92 & 267.07 &  89.02 \\
\hline
14.377 & 0.95 & 150$^\circ$ & 2031.41 & 1354.27 & 1127.47 & 563.74 & 273.04 &  91.01 \\
\hline
14.663 & 0.95 & 170$^\circ$ & 2068.15 & 1378.77 & 1147.02 & 573.51 & 275.07 &  91.69 \\
\hline
\end{tabular}
\end{ruledtabular}
\caption{Same as Table {\ref{tab:dura0.1}}, but for $a = 0.4M$.  As in
  the $a = 0.1M$ case, the number of oscillations spent near each
  resonance is quite large.}
\label{tab:dura0.4}
\end{table*}

\begin{table*}[ht]
\begin{ruledtabular}
\begin{tabular}{|c|c|c|c|c|c|c|c|c|}
\hline
$p_{\rm init}/M$ & $e_{\rm init}$ & $\theta_{\rm inc,\ init}$
& $N_{\theta,3:2}$ & $N_{r,3:2}$ & $N_{\theta,2:1}$ & $N_{r,2:1}$ 
& $N_{\theta,3:1}$ & $N_{r,3:1}$ \\
\hline
7.304 & 0.1 & 10$^\circ$ & 1250.91 & 833.94 & 610.35 & 305.18 & 208.82 &  69.61 \\
\hline
7.832 & 0.1 & 30$^\circ$ & 1331.61 & 887.74 & 656.36 & 328.18 & 225.65 &  75.22 \\
\hline
9.460 & 0.1 & 60$^\circ$ & 1575.42 & 1050.28 & 792.41 & 396.21 & 274.98 &  91.66 \\
\hline
11.253 & 0.1 & 85$^\circ$ & 1845.88 & 1230.59 & 941.51 & 470.76 & 328.24 & 109.41 \\
\hline
12.001 & 0.1 & 95$^\circ$ & 1960.00 & 1306.66 & 1004.36 & 502.18 & 350.79 & 116.93 \\
\hline
13.728 & 0.1 & 120$^\circ$ & 2226.87 & 1484.58 & 1151.47 & 575.74 & 403.04 & 134.35 \\
\hline
15.224 & 0.1 & 150$^\circ$ & 2461.90 & 1641.26 & 1281.02 & 640.51 & 448.97 & 149.66 \\
\hline
15.708 & 0.1 & 170$^\circ$ & 2537.83 & 1691.89 & 1322.86 & 661.43 & 463.78 & 154.59 \\
\hline
\hline
7.381 & 0.4 & 10$^\circ$ & 1248.69 & 832.46 & 623.44 & 311.72 & 212.42 &  70.81 \\
\hline
7.909 & 0.4 & 30$^\circ$ & 1320.58 & 880.38 & 668.19 & 334.09 & 229.00 &  76.33 \\
\hline
9.526 & 0.4 & 60$^\circ$ & 1545.63 & 1030.42 & 800.30 & 400.15 & 275.55 &  91.85 \\
\hline
11.319 & 0.4 & 85$^\circ$ & 1799.60 & 1199.73 & 942.92 & 471.46 & 320.13 & 106.71 \\
\hline
12.078 & 0.4 & 95$^\circ$ & 1918.26 & 1278.84 & 1009.92 & 504.96 & 343.97 & 114.66 \\
\hline
13.816 & 0.4 & 120$^\circ$ & 2173.66 & 1449.10 & 1151.51 & 575.75 & 384.48 & 128.16 \\
\hline
15.323 & 0.4 & 150$^\circ$ & 2402.57 & 1601.72 & 1277.81 & 638.90 & 419.76 & 139.92 \\
\hline
15.807 & 0.4 & 170$^\circ$ & 2476.78 & 1651.18 & 1318.61 & 659.31 & 430.88 & 143.63 \\
\hline
\hline
7.546 & 0.7 & 10$^\circ$ & 1249.07 & 832.71 & 661.42 & 330.71 & 224.99 &  75.00 \\
\hline
8.063 & 0.7 & 30$^\circ$ & 1306.00 & 870.67 & 700.63 & 350.31 & 240.21 &  80.07 \\
\hline
9.669 & 0.7 & 60$^\circ$ & 1499.00 & 999.33 & 817.77 & 408.88 & 275.58 &  91.86 \\
\hline
11.473 & 0.7 & 85$^\circ$ & 1724.10 & 1149.40 & 940.78 & 470.39 & 295.16 &  98.39 \\
\hline
12.232 & 0.7 & 95$^\circ$ & 1848.31 & 1232.20 & 1018.83 & 509.42 & 324.80 & 108.27 \\
\hline
13.992 & 0.7 & 120$^\circ$ & 2082.42 & 1388.28 & 1145.27 & 572.63 & 341.14 & 113.71 \\
\hline
15.521 & 0.7 & 150$^\circ$ & 2298.83 & 1532.55 & 1262.88 & 631.44 & 359.29 & 119.76 \\
\hline
16.005 & 0.7 & 170$^\circ$ & 2369.21 & 1579.47 & 1300.96 & 650.48 & 364.98 & 121.66 \\
\hline
\hline
7.744 & 0.95 & 10$^\circ$ & 1263.91 & 842.61 & 720.75 & 360.37 & 259.37 &  86.46 \\
\hline
8.250 & 0.95 & 30$^\circ$ & 1308.75 & 872.50 & 743.76 & 371.88 & 263.68 &  87.89 \\
\hline
9.845 & 0.95 & 60$^\circ$ & 1473.91 & 982.61 & 831.54 & 415.77 & 263.92 &  87.97 \\
\hline
11.660 & 0.95 & 85$^\circ$ & 1672.12 & 1114.74 & 926.52 & 463.26 & 251.94 &  83.98 \\
\hline
12.419 & 0.95 & 95$^\circ$ & 1797.82 & 1198.55 & 1019.35 & 509.68 & 290.65 &  96.88 \\
\hline
14.201 & 0.95 & 120$^\circ$ & 2010.42 & 1340.28 & 1125.18 & 562.59 & 286.82 &  95.61 \\
\hline
15.752 & 0.95 & 150$^\circ$ & 2212.45 & 1474.97 & 1231.09 & 615.55 & 294.38 &  98.13 \\
\hline
16.247 & 0.95 & 170$^\circ$ & 2278.52 & 1519.02 & 1265.56 & 632.78 & 297.07 &  99.02 \\
\hline
\end{tabular}
\end{ruledtabular}
\caption{Same as Table {\ref{tab:dura0.1}}, but for $a = 0.7M$.  Again we
see many oscillations near each resonance.}
\label{tab:dura0.7}
\end{table*}

\begin{table*}[ht]
\begin{ruledtabular}
\begin{tabular}{|c|c|c|c|c|c|c|c|c|}
\hline
$p_{\rm init}/M$ & $e_{\rm init}$ & $\theta_{\rm inc,\ init}$
& $N_{\theta,3:2}$ & $N_{r,3:2}$ & $N_{\theta,2:1}$ & $N_{r,2:1}$ 
& $N_{\theta,3:1}$ & $N_{r,3:1}$ \\
\hline
5.555 & 0.1 & 10$^\circ$ & 1040.54 & 693.69 & 496.24 & 248.12 & 171.77 &  57.26 \\
\hline
6.303 & 0.1 & 30$^\circ$ & 1146.72 & 764.48 & 556.67 & 278.34 & 192.09 &  64.03 \\
\hline
8.547 & 0.1 & 60$^\circ$ & 1465.33 & 976.89 & 733.84 & 366.92 & 255.86 &  85.29 \\
\hline
10.956 & 0.1 & 85$^\circ$ & 1816.70 & 1211.14 & 926.27 & 463.14 & 323.75 & 107.92 \\
\hline
11.946 & 0.1 & 95$^\circ$ & 1964.41 & 1309.61 & 1007.32 & 503.66 & 352.56 & 117.52 \\
\hline
14.190 & 0.1 & 120$^\circ$ & 2308.88 & 1539.26 & 1196.90 & 598.45 & 419.66 & 139.89 \\
\hline
16.115 & 0.1 & 150$^\circ$ & 2611.50 & 1741.00 & 1363.62 & 681.81 & 478.63 & 159.54 \\
\hline
16.731 & 0.1 & 170$^\circ$ & 2709.19 & 1806.12 & 1417.44 & 708.72 & 497.64 & 165.88 \\
\hline
\hline
5.621 & 0.4 & 10$^\circ$ & 1059.72 & 706.48 & 504.33 & 252.16 & 172.90 &  57.63 \\
\hline
6.358 & 0.4 & 30$^\circ$ & 1145.20 & 763.47 & 565.87 & 282.93 & 194.22 &  64.74 \\
\hline
8.602 & 0.4 & 60$^\circ$ & 1432.90 & 955.27 & 741.04 & 370.52 & 259.49 &  86.50 \\
\hline
11.011 & 0.4 & 85$^\circ$ & 1764.06 & 1176.04 & 926.30 & 463.15 & 319.71 & 106.57 \\
\hline
12.001 & 0.4 & 95$^\circ$ & 1918.03 & 1278.69 & 1012.30 & 506.15 & 349.15 & 116.38 \\
\hline
14.267 & 0.4 & 120$^\circ$ & 2249.57 & 1499.72 & 1195.52 & 597.76 & 401.35 & 133.78 \\
\hline
16.203 & 0.4 & 150$^\circ$ & 2545.56 & 1697.04 & 1358.24 & 679.12 & 445.64 & 148.55 \\
\hline
16.830 & 0.4 & 170$^\circ$ & 2641.47 & 1760.98 & 1410.78 & 705.39 & 459.60 & 153.20 \\
\hline
\hline
5.753 & 0.7 & 10$^\circ$ & 1098.79 & 732.53 & 531.67 & 265.84 & 176.33 &  58.78 \\
\hline
6.479 & 0.7 & 30$^\circ$ & 1147.10 & 764.73 & 596.76 & 298.38 & 201.32 &  67.11 \\
\hline
8.701 & 0.7 & 60$^\circ$ & 1382.96 & 921.97 & 759.38 & 379.69 & 271.07 &  90.36 \\
\hline
11.132 & 0.7 & 85$^\circ$ & 1678.84 & 1119.23 & 921.04 & 460.52 & 303.58 & 101.19 \\
\hline
12.133 & 0.7 & 95$^\circ$ & 1840.54 & 1227.03 & 1020.42 & 510.21 & 337.85 & 112.62 \\
\hline
14.421 & 0.7 & 120$^\circ$ & 2147.35 & 1431.57 & 1185.83 & 592.92 & 357.66 & 119.22 \\
\hline
16.401 & 0.7 & 150$^\circ$ & 2429.46 & 1619.64 & 1337.90 & 668.95 & 378.86 & 126.29 \\
\hline
17.028 & 0.7 & 170$^\circ$ & 2521.06 & 1680.71 & 1387.04 & 693.52 & 385.64 & 128.55 \\
\hline
\hline
5.918 & 0.95 & 10$^\circ$ & 1144.25 & 762.83 & 617.07 & 308.54 & 184.00 &  61.33 \\
\hline
6.622 & 0.95 & 30$^\circ$ & 1165.19 & 776.80 & 672.55 & 336.28 & 219.48 &  73.16 \\
\hline
8.833 & 0.95 & 60$^\circ$ & 1358.49 & 905.66 & 779.79 & 389.89 & 292.49 &  97.50 \\
\hline
11.275 & 0.95 & 85$^\circ$ & 1620.62 & 1080.41 & 903.79 & 451.89 & 266.61 &  88.87 \\
\hline
12.287 & 0.95 & 95$^\circ$ & 1784.38 & 1189.59 & 1021.31 & 510.65 & 312.28 & 104.09 \\
\hline
14.619 & 0.95 & 120$^\circ$ & 2065.73 & 1377.16 & 1161.82 & 580.91 & 302.29 & 100.76 \\
\hline
16.632 & 0.95 & 150$^\circ$ & 2330.75 & 1553.83 & 1299.25 & 649.62 & 309.27 & 103.09 \\
\hline
17.270 & 0.95 & 170$^\circ$ & 2417.14 & 1611.43 & 1343.78 & 671.89 & 312.11 & 104.04 \\
\hline
\end{tabular}
\end{ruledtabular}
\caption{Same as Table {\ref{tab:dura0.1}}, but for $a = 0.9M$.  Again
  we see many oscillations near each resonance.}
\label{tab:dura0.9}
\end{table*}

\end{document}